\documentclass[10pt]{article}

\usepackage{amsmath}
\usepackage{amssymb}

\usepackage{graphicx}

\usepackage{cite}

\usepackage{color}

\usepackage{longtable}
\usepackage[labelfont=bf,labelsep=period,justification=raggedright]{caption}

\makeatletter
\renewcommand{\@biblabel}[1]{\quad#1.}
\makeatother

\date{}

\begin{document}

\begin{flushleft}
{\Large
\textbf{Low-dimensional clustering detects incipient
dominant influenza strain clusters}
}
\\

Jiankui He$^{1}$ and Michael W. Deem$^{1,2}$
\\
 $^1$Department of Physics \& Astronomy, Rice University, Houston, Texas, USA
\\
$^2$Department of Bioengineering, Rice University, Houston, Texas, USA
\\
$\ast$ E-mail: Corresponding mwdeem@rice.edu
\end{flushleft}

\section*{Abstract}

Influenza has been circulating in the human population and has
caused three pandemics in the last century (1918 H1N1, 1957 H2N2,
1968 H3N2). The 2009 A(H1N1) was classified by
the World Health Organization (WHO) as the fourth pandemic.
Influenza has a high evolution rate, which makes vaccine design
challenging.
We here consider an approach
for early detection of new dominant strains.
By clustering the 2009 A(H1N1) sequence data, we found two main clusters. We then
define a metric to detect the emergence of dominant strains.  We
show on historical H3N2 data that this method is able to identify a
cluster around an incipient dominant strain before it becomes
dominant. For example, for H3N2 as of March 30, 2009, the method
detects the
cluster for the new A/British Columbia/RV1222/2009 strain.
This strain detection tool would appear to be useful for annual influenza
vaccine selection.\\
 \emph{Keywords:} clustering/H1N1/H3N2/influenza 

\section*{Introduction}
The recent outbreak of 2009 A(H1N1) caused immediate international
attention\cite{deem2009,Garten2009,Fraser2009,smith2009}. This  new
2009 A(H1N1) virus contains a combination of gene segments from
swine and human influenza viruses\cite{Garten2009,Fraser2009}.
Confirmed infections  reached 270,000
globally as of September 2009\cite{who0501}.  The novel 2009 A(H1N1) strain was defined
as a pandemic strain  by the World health Organization(WHO) in 2009\cite{whopandemic}, and was the epidemic
strain in the 2009 Northern winter.

Influenza viruses are hyper-mutating viruses. It has been estimated that
the nucleotide mutation rate per genome per replication is approximately 0.76\cite{drake1999}.
Influenza viruses escape the
 human immune system by continual antigenic
drift and shift\cite{fitch1997, webster1998,gupta1998,ferguson2003,ghedin2005,Nelson2007}.
The quasispecies nature of influenza viruses makes the strain structure complex\cite{domingo2002}.  Usually, there is one or a few dominant influenza strains
circulating in the population for each flu season. The flu vaccine
is most effective when it matches this dominant circulating
strain\cite{hak2002,gupta2006}. The degree to which immunity induced by a
vaccine
 protects against a
different viral strain is determined by the antigenic distance
between the vaccine and the virus.
 Due to evolution of the antigenic
regions of the influenza virus, the composition of the flu vaccine
is typically modified annually\cite{Russell2008vaccine}. However, since the influenza strains
used in the flu vaccine are decided 6 months before the flu season,
a mismatch between the vaccine strain and dominant circulating
strain may occur if the virus evolves significantly.
 Such a
situation arose for the H3N2 virus in the 2009--2010 flu season, when
 A/British Columbia/RV1222/2009 emerged in the early spring\cite{promed1,promed2}. Accurate early prediction of the dominant circulating
strain is an essential and important task in influenza research.

Understanding the evolution of influenza viruses has benefited from phylogenetic reconstructions of the hemagglutinin protein
evolution\cite{Russell2008,ferguson2003}.  In an
alternative approach, Lapedes and Farber\cite{lapedes2001}, followed by
Smith \emph{et al.}\cite{Smith2004}, applied a technique called multidimensional scaling to
study antigenic evolution of influenza.
Plotkin \emph{et al.} clustered hemagglutinin protein sequences using the
single-linkage clustering algorithm and found that
influenza viruses group into clusters\cite{Joshua2002}.

Here, we present a low-dimensional clustering method that can
 detect the  cluster containing
an incipient dominant strain for an upcoming flu season before
the strain becomes dominant. The method builds upon the dimensional
projection technique used by
Lapedes and Farber\cite{lapedes2001} and
Smith \textit{et al.}\cite{Smith2004}
to characterize hemagglutination inhibition data. In this paper, we first study the evolution of 2009 A(H1N1) by an evolutionary path map
which leads to a suggestion for
the H1N1 vaccine strain.
Then, we introduce the low-dimensional protein sequence clustering method.
We propose an influenza vaccine selection procedure
based on this sequence clustering. The procedure
is demonstrated and tested in detail using historical data.
We show the performance of the method to
predict the dominant H3N2 strain in an upcoming flu season using
data solely from before the flu season, on
data since 1996.
We  compare the results
to those from existing methods since 1996.
In the discussion section, we discuss the
relationship between the protein sequence clustering method and
previous approaches.
We discuss the false positive rate, as well as other challenges.

\section*{Results}
\subsection*{Evolutionary path of 2009 A(H1N1) influenza}

\begin{figure}
\begin{center}
\includegraphics[height=4in]{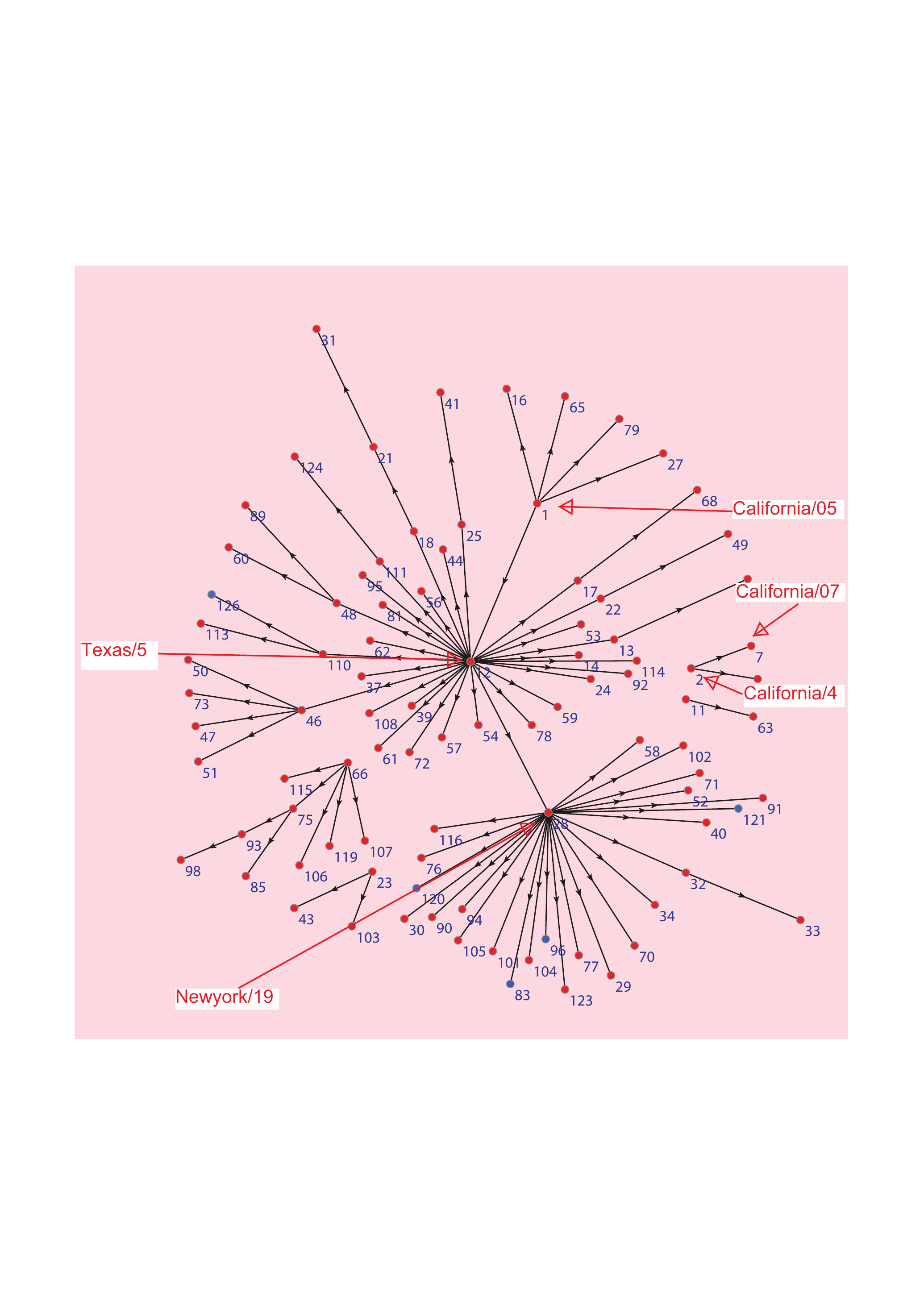}\\
\caption{The evolutionary path of 2009 A(H1N1) influenza.
Strain \#1: A/California/05/2009. Strain \#2: A/California/04/2009. Strain
\#7: A/California/07/2009. Strain \#12: A/Texas/05/2009. Strain
\#28: A/New York/19/2009. For complete strain names, see
supplemental material. Strains from the Northern and Southern
hemisphere are shown as red dots and blue dots respectively. One branch represents one substitution in the amino acid sequence.
  \label{path} }
\end{center}
\end{figure}

We first construct the directional evolutionary path for the 2009
A(H1N1) influenza. We use high resolution data in sequence, time,
and world spatial coordinate to construct this evolutionary
relationship. Since its first detection, the 2009 A(H1N1) virus has
been extensively sequenced\cite{Garten2009,Fraser2009}. By
May 1, 2009, the number of confirmed cases reported by
WHO was 333\cite{who0501}. At the same time, the sequenced
hemagglutinin protein (HA) available in NCBI Influenza Resources
Database were 312\cite{bao2008}; that is to say most of the
confirmed cases at that time were sequenced. At July 1, 2009, the
ratio of sequenced HA protein to confirmed cases by WHO was $1039 / 77201$\cite{who0501}, a number which is still much
larger than that for seasonal flu. In addition, the Influenza
Resources Database contains the date of collection of each 2009 A(H1N1)
virus strain. We
reconstruct the evolutionary history of swine flu viruses
with the following procedure.
If strain B is mutated from strain A, we term strain A ``founder'' and strain B ``F1''
 We align the HA proteins of all 2009 A(H1N1) strains. Then,
for each strain, we find its  founder strain based on the following
four criteria: 1, the  founder strain should appear earlier than the
strain, as judged by collection date;
2, the  founder strain should have only one
amino acid difference in the HA1 protein relative to the F1 strain; 3, the
 founder should also have the most similar nucleotide sequence
relative to F1;  and 4, the founder strain should have a large
number of identical copies circulating in human population, as
approximated by the number of different strains with identical HA
sequences in the Influenza Resources Database. By applying these
four criteria to 2009 A(H1N1) influenza, we construct the directional
evolutionary path map, as shown in Fig.\ \ref{path}. We can
see two clusters: one around A/New York/19/2009 (\#28), and
another one around A/Texas/05/2009 (\#12). Most new strains are
from the Northern hemisphere, and strains from the Southern
hemisphere are mainly located at the edge of the map, such as strain
\#96, \#120, and \#126. Geographically, we see many founder to F1
 links are from US and Mexico to other countries, but we
rarely see founder to F1 links that are from other countries
to US and Mexico, or from other countries to other countries except
US and Mexico (see Materials and Methods). We also found that
strains with more F1 in Fig.\ \ref{path} are more frequently
seen in the human population. For example, in the Influenza
Resources Database, we found 153 strains to be identical with A/New
York/19/2009, which has 29 F1 strains, and 120 strains to be
identical with A/Texas/05/2009, which has 24 F1 strains. We
can see in Fig.\ \ref{path} that A/Texas/05/2009 is at the very
upstream of the map, with downward connections to most of the other
strains by direct or two-step links.  This result agrees with the US Food
and Drug Administration\cite{fda} recommendation of A/Texas/05/2009 as
a vaccination strain. The alternative vaccine strain A/California/7/2009 (\#7) has fewer F1 strains and it
is not located at the center of the network.

\subsection*{Low-dimensional clustering}
\begin{figure}
\begin{center}
\includegraphics[width = 0.43\textwidth]{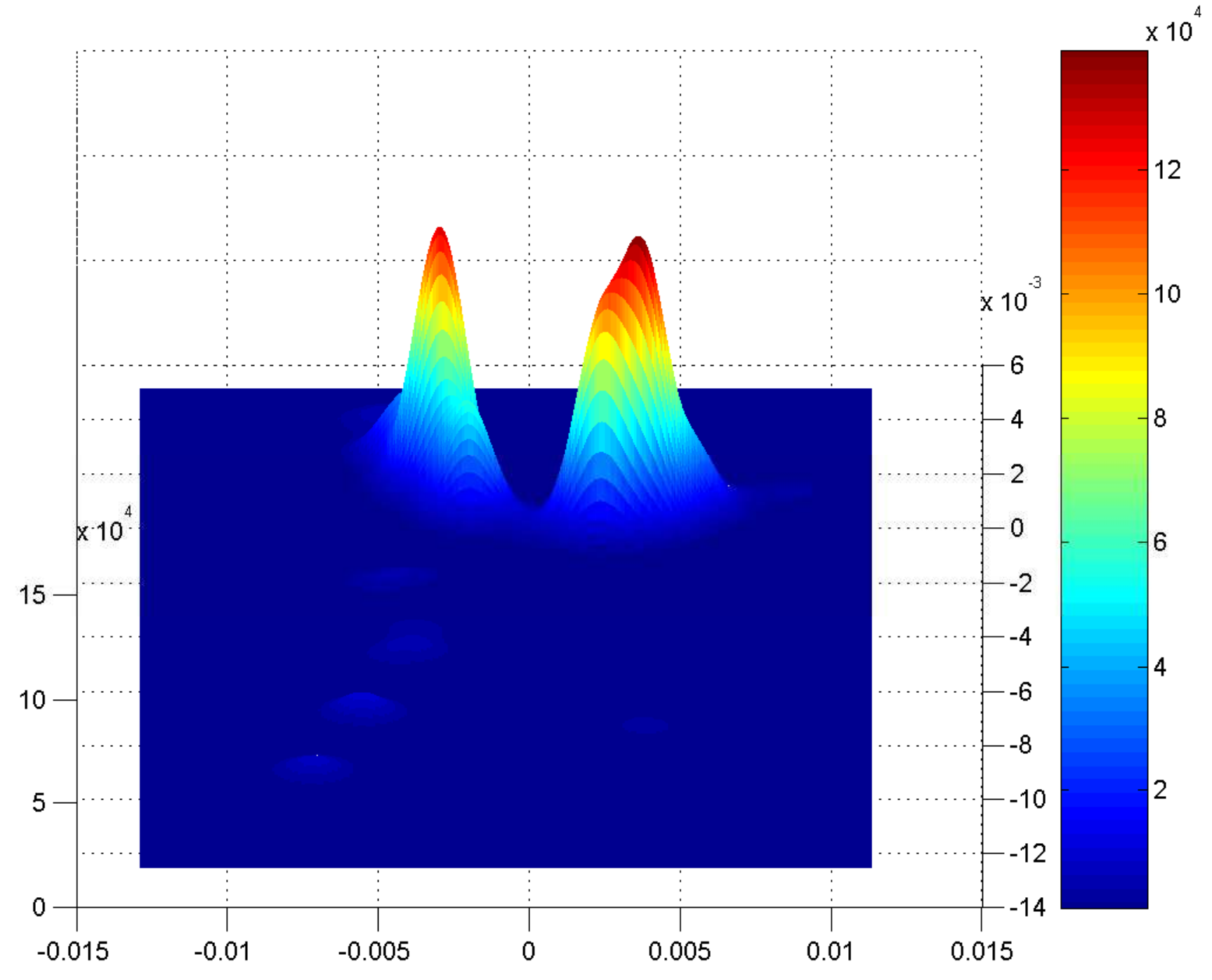}
\includegraphics[width = 0.43\textwidth]{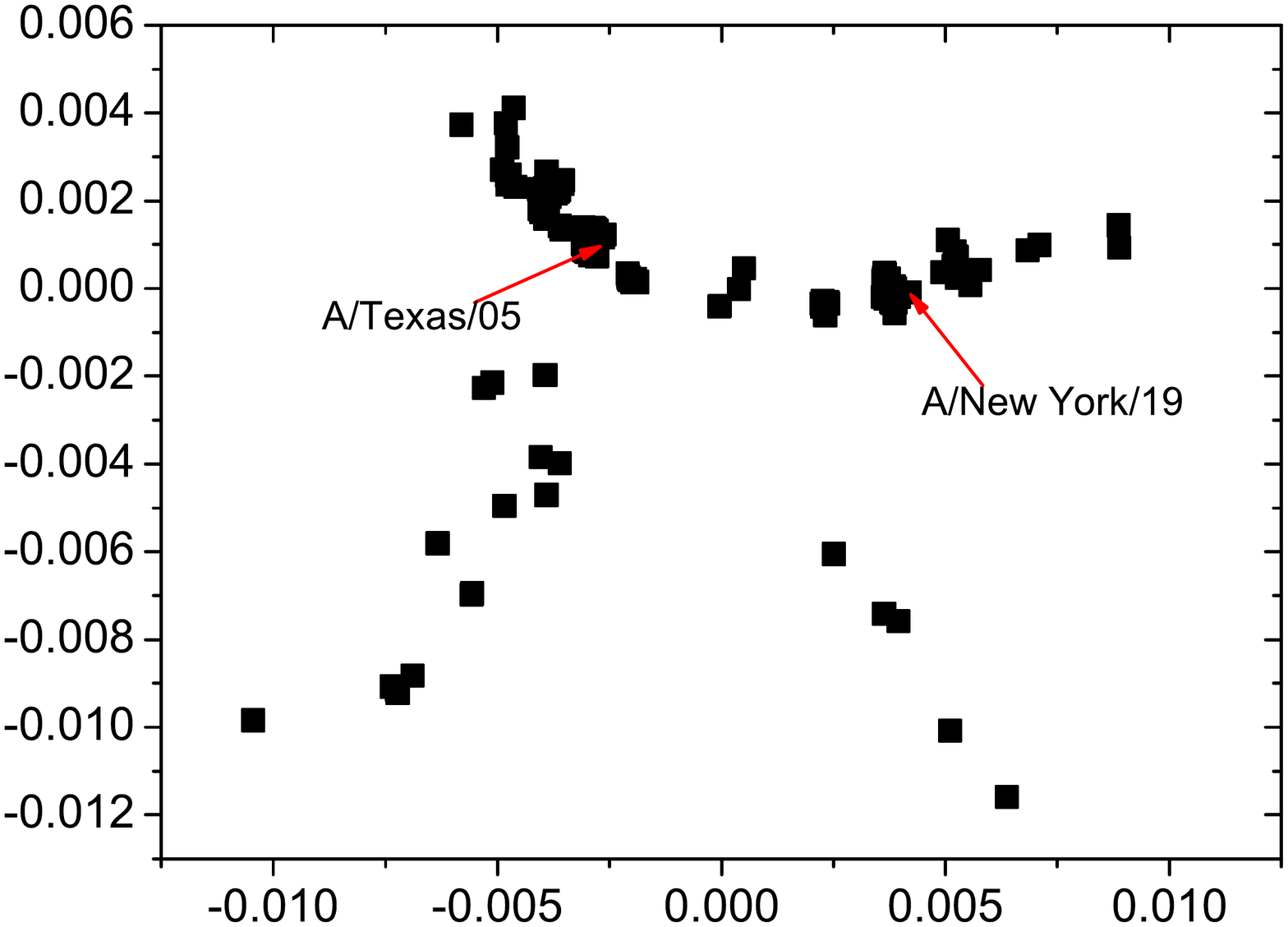}\\
(a)  \hspace{2.4in}  (b)\\
\caption{(a), Kernel density estimation for the protein distance map
of 2009 A(H1N1) influenza as of December 5, 2009. (b), The protein distance map of 2009
A(H1N1) influenza. The vertical and horizontal axes of both figures represent protein distance as defined in Equation \ref{eq}. A 0.0018 unit of
protein distance equals one substitution in the HA protein sequence of H1N1. The height and colors in (a) both represent the density of isolates.
  \label{map_H1N1} }
\end{center}
\end{figure}

We use a low-dimensional clustering method to visualize the
antigenic distance matrix of the viruses. We use a statistical tool
called ``multidimensional scaling"\cite{everitt2001}.  This
 method was used by
Lapedes and Farber\cite{lapedes2001} and
Smith \textit{et al.}\cite{Smith2004}
to project
ferret hemagglutination inhibition assay data to low
dimensions. The influenza viral surface glycoprotein hemagglutinin is a
 primary target of the protective immune response. Here we project the hemagglutinin protein sequence data, rather than animal model data, to low dimensions.
The HA1 protein of influenza with 329 residues can be considered as a
329-dimension space. The multidimensional scaling method is applied
to rescale the 329-dimension space to a 2-dimensional space, so that
we can plot and visualize it. First, we do a multialignment of the HA1
proteins. Then, the distance between any two proteins is calculated
as
\begin{equation}
d_{ij}=\frac{1}{N}\sum_{m=1}^N (1-\delta_{s_{i,m},s_{j,m}})
\label{eq}
\end{equation}
where $s_{i,m}$ is the amino acid of protein $i$ at position $m$.  The term $\delta_{s_{i,m},s_{j,m}}$ is 1 if amino acids of protein $i$ and $j$ at position $m$ are the same. Otherwise, it is 0. For the 2009 H1N1 viruses, we consider the entire HA protein, and $N=566$. For H3N2 viruses, we consider only the HA1 protein,  and $N=329$, because the entire HA proteins are not completely sequenced in many cases. Thus, $d_{ij}$ is the number of amino acid differences between HA proteins normalized by length. The multidimensional scaling produces a protein distance map, for example, Fig.\ \ref{map_H1N1}(b).
In this map, each data point represents a flu strain isolate. The Euclidean distance between two points in the map
approximates the protein distance in Equation \ref{eq} between these two flu strains (see Materials and Methods for details of this distance approximation
procedure). Two closely located points imply two
strains with similar HA protein sequences.

We apply the low-dimensional clustering method to study 2009
A(H1N1). We plot the protein distance map in Fig.\
\ref{map_H1N1}(b). Both A/Texas/05/2009 and A/New York/19/2009 are
located near the center of the cluster, in good agreement with the
observation from Fig.\ \ref{path} that they are the founder strains
for many F1 strains. To detect the clusters in the protein
distance map, we use a statistical method known as kernel density
estimation\cite{everitt2001}. Kernel density estimation  is a
non-parametric method to estimate the probability density function
from which data come. The kernel density figure is produced from
the protein distance map, and it shows the density of influenza
strains in
sequence space. We plot the kernel density as the three dimensional shaded surface.
For example, the kernel density surface Fig. \ref{map_H1N1}(a) is produced from Fig.\
\ref{map_H1N1}(b). The $x$ and $y$ axes in Fig.\ \ref{map_H1N1}(a) are the same as that in Fig.\ \ref{map_H1N1}(b) and are protein distance coordinates. The $z$ dimension measures the density of flu strains around point $(x,y)$. We use the surface height and the colors to represent $z$ values, and the color is proportional to surface height. A peak in kernel density Fig.\ \ref{map_H1N1}(a) indicates a cluster of related flu strains in the protein distance map Fig.\ \ref{map_H1N1}(b)

There are two significant clusters in the Fig.\ \ref{map_H1N1}(a), as two peaks are observed.
The cluster on the left side contains A/Texas/05/2009. Another cluster on the right side
contains A/New York/19/2009.  The 2009 A(H1N1) virus
has evolved slowly to date. The greatest \textit{p}$_\textrm{epitope}$
antigenic distance (definition in the Materials and Methods)
between A/Texas/05/2009 and all sequenced  strains
is measured to be
 $<0.08$. Values of
\textit{p}$_\textrm{epitope}$ less than 0.45 for H1N1 indicate
positive expected vaccine efficacy\cite{keyao2009}, and so a vaccine is
expected to be efficacious.  All of the amino acids in all five epitopes of a strain of A/Texas/05/2009
and a strain of A/New York/19/2009 are the same. Multidimensional scaling
predicts that A/Texas/05/2009 will be the dominant strain in the
2009--2010 season, and that A/Texas/05/2009 is a suitable strain for
vaccination. Our focus is on the expected vaccine effectiveness, as it can be judged from antisera HI assay or sequence data alone. We do not consider other aspects such as growth in hen's eggs or other manufacturing constraints. Laboratory growth and passage data are needed to address these aspects.

\subsection*{H3N2 virus evolution for 40 years}

\begin{figure}
\begin{center}
\includegraphics[width = 0.63\textwidth]{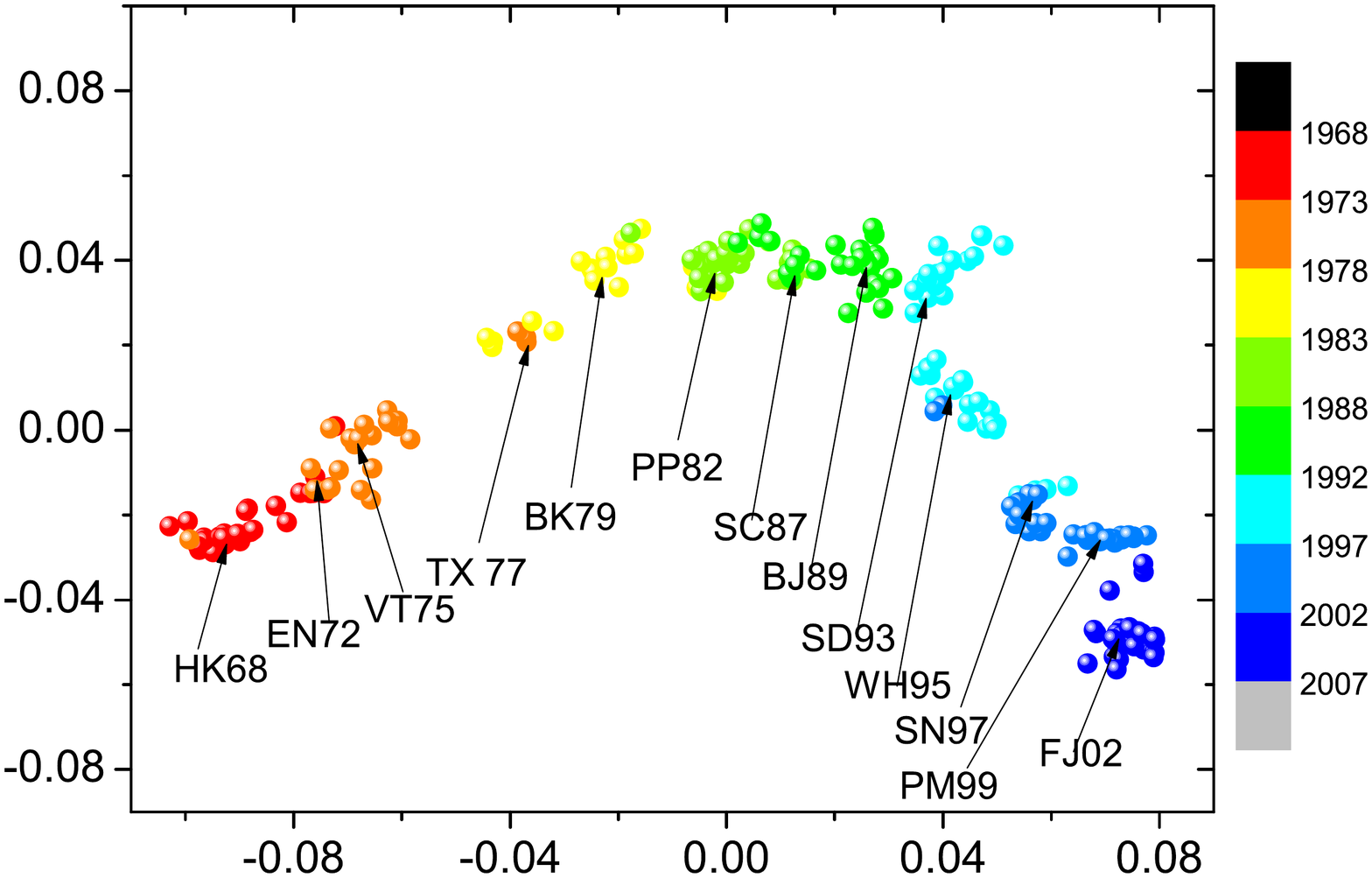}\\
(a)\\
\includegraphics[width=0.73\textwidth]{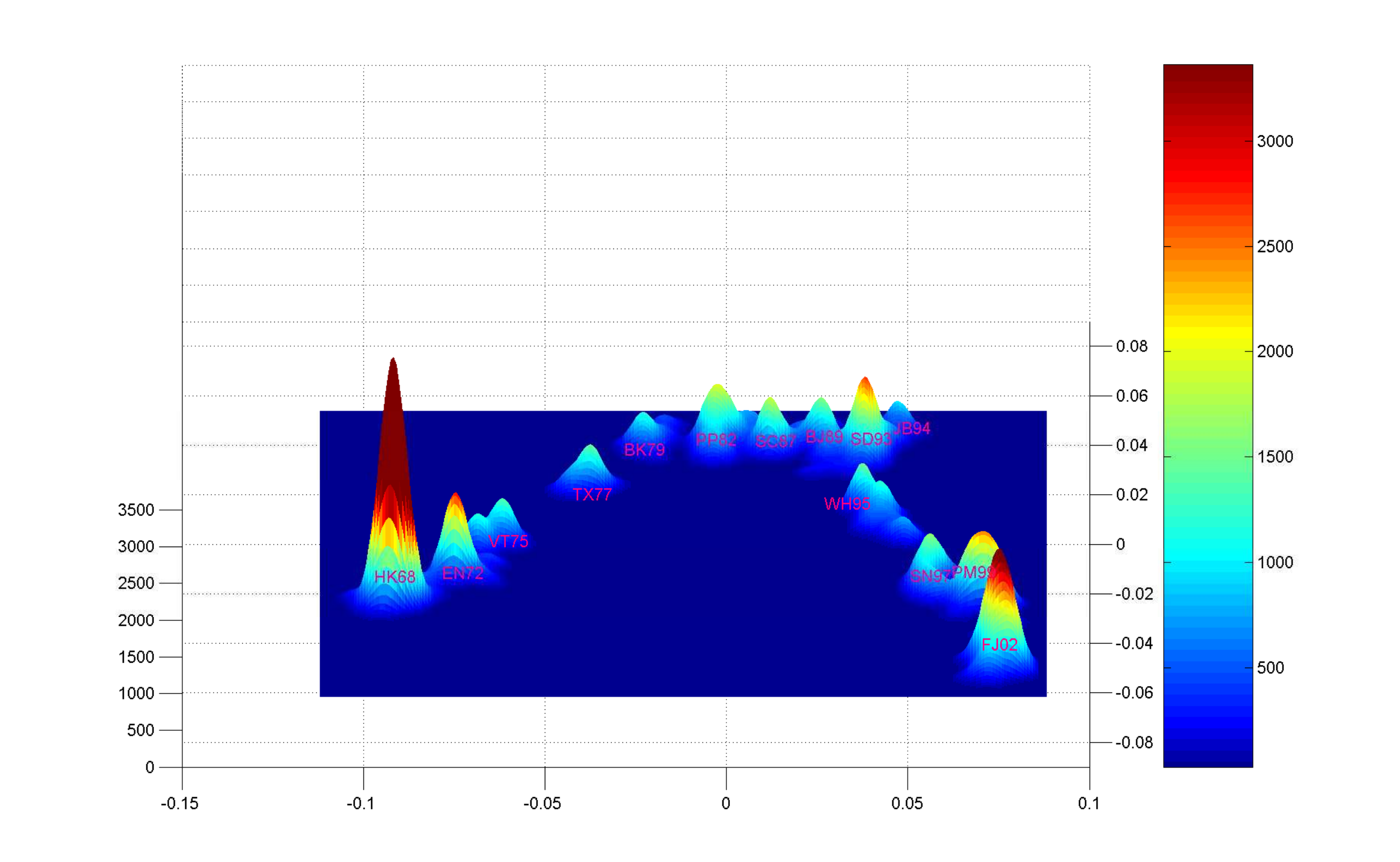}\\
  (b)\\
\caption{(a) The protein distance map and (b) corresponding Kernel density estimation
of influenza from 1968 to 2007.  The vertical and horizontal axes of both figures represent protein distance as defined in Equation \ref{eq}. A 0.0030 unit of protein distance equals one substitution in the HA1 protein sequence of H3N2. The colors in (a) represent the time of collection of the isolates. The colors and height in (b) represent the density of isolates. Each cluster is named after the first vaccine strain in the cluster.  HK68: Hongkong/1/68, EN72: England/42/72, VT75: Victoria/3/75, TX77: Texas/1/77, BK79: Bangkok/1/79, PP82: Philippines/2/82, SC87: Sichuan/2/87, BJ89: Beijing/32/92, SD93: Shandong/9/93, JB94: Johannesburg/33/94, WH95: Wuhan359/95, SN97: Sydney/5/97, PM99: Panama/2007/99, FJ02: Fujian/411/2002;
  \label{map_allyears} }
\end{center}
\end{figure}

We construct the protein distance map to determine the evolution of
influenza A(H3N2) virus from 1969 to 2007. Sequences of HA1 proteins were downloaded
from the Influenza Virus Resources database\cite{bao2008}.
We use the multidimensional clustering method\cite{lapedes2001} to generate the protein distance map and
corresponding kernel density estimation in Fig.\ \ref{map_allyears}.
Smith \emph{et al.}\cite{Smith2004} produced a similar graph using ferret antisera HI assay data. The figure presented
here has a higher resolution, and more clusters are observed, because protein sequences data are more abundant and accurate than antisera HI assay data.
The evolution of influenza tends to group strain into clusters.
In Fig.\ \ref{map_allyears}, we identified 14 major clusters
by setting a cutoff value of kernel density. We marked each cluster by the first vaccine strain in the cluster.
 The average duration time for a cluster is 2.7 years, which is also the approximate duration of a vaccine.
 There are apparent gaps between clusters.  The antigenic distance between two strains in two separate  clusters is larger than the distances within the same cluster. The influenza virus evolves within one cluster before jumping from one cluster to another cluster. This dynamics occurs because small antigenic drift by one or a few sequential mutations does not lead the virus to completely escape from  cross immunity induced by vaccine protection or prior exposure.

For vaccine design, when the viruses evolve as a quasispecies in the same cluster, the vaccine that is targeted to the cluster provides protection. This protection decreases with antigenic distance. When the viruses jump to a new cluster by antigenic drift or shift, one would want to update the vaccine to provide protection against strains in the new cluster. In Fig.\ \ref{map_allyears}(a), the arrows point to the exact position of vaccine strains. It can be seen that the positions of vaccine strains are  near the center of clusters. It can be shown mathematically that choosing the consensus strain of a cluster as vaccine strain minimizes the \textit{p}$_\textrm{epitope}$  antigenic distance between vaccine strain and cluster strains, and thus maximizes expected vaccine efficacy.

\subsection*{Influenza vaccine strain selection}

We now use the low-dimensional sequence clustering method in an effort to
 detect a new flu strain before it becomes dominant. A
question of interest in the influenza research is whether we can
predict which strain will be dominant in the next flu season based
on the information we have at present. The WHO gathers together
every February to make a recommendation for influenza strains to be
used in vaccine for next flu season in the Northern hemisphere.
The vaccine
is expected to have high efficacy if the chosen strain is dominant in the next flu
season. The recommendation is especially challenging to make when the dominant strain in next flu season has not been dominant
before February of that year. For example, in mid-March 2009, a new H3N2 strain appeared\cite{promed1,promed2}, which infected a significant fraction of the population in the Southern hemisphere.

The current accepted influenza vaccine strain selection procedure is as follows \cite{Russell2008vaccine}.
Isolates samples are collected by WHO GISN and are characterized
antigenically using the hemagglutination inhibition(HI) assay. About
10\% of samples are also sequenced in HA1 domain of HA gene. Antigenic maps are constructed from the HI assay data using dimensional projection technique. Examination of HI data is not dependent on analysis using dimensional projection, but rather, the primary HI data may carry the most weight.
  If the vaccine does not match the current circulating strains, the vaccine is updated to contain one representative of the circulating strains. The emerging variant strains are identified. If the antigenically distinct emerging variants are judged to be the dominant strains in the upcoming season, the vaccine is updated to include one representative of emerging variants. The key issue and major difficulty is how to judge whether emerging variants will be the dominant variants in next season. If a fourfold difference in antisera HI titer between the vaccine strain and the emerging strains is observed, the emerging strain is to be determined to be dominant strains in upcoming season, and an updated vaccine is recommended to include the emerging strains\cite{Russell2008vaccine}.

 Here, we propose a modified vaccine selection process based on clustering detection.
 First, we apply the multidimensional scaling to
make a protein distance map from HA1 sequences, instead of constructing an antigenic map from HI assay data.
Then, we use kernel density estimation to determine the clusters of strains.
If the vaccine does not match the current circulating cluster, the vaccine is updated to contain the current circulating strain. If the vaccine matches the current circulating cluster, but an emerging cluster is judged likely to
be the major cluster in the upcoming season, the vaccine is updated to contain the consensus strain of the emerging cluster.  We judge whether a cluster is an emerging dominant cluster by
two criteria. The first criterion is that this cluster can be  detected by kernel density
estimation, and is separate from the cluster that contains the
current circulating strain or vaccine strain. A cluster that can be detected by kernel density
estimation usually contains a central strain that has multiple identical copies and some F1 strains
that are closely related to the central strain.  An example is the cluster of A/Texas/05/2009(H1N1) in Fig.\ \ref{path}.  A/Texas/05/2009(H1N1) is the central strain, which has 120 strains with identical HA protein sequences in the Influenza Virus Resource database\cite{bao2008}. A/Texas/05/2009(H1N1) also has 29 F1 strains with one amino acid different. So, A/Texas/05/2009(H1N1) and the surrounding strains form a cluster as we detected in Fig.\ \ref{map_H1N1} by kernel density estimation.

 The second criterion is that the current vaccine strain does not match the consensus strain of the cluster and is estimated to provide low protection against strains in the cluster. That is, an immune response stimulated by a vaccine cannot effectively protect against infection by sufficiently distant by new strains. The consensus strain is a protein sequence that shows which residues are most abundant in the multialignment at each position. The efficacy of current vaccine to the new cluster can be estimated from ferret antisera HI assay data. However, the antisera data has low resolution and has an imperfect correlation to vaccine effectiveness in humans\cite{zhou2010,gupta2006}. Instead, we use \textit{p}$_\textrm{epitope}$, which is calculated as the fraction of mutations in dominant epitope, to estimate vaccine efficacy and which has a more robust correlation to vaccine effectiveness in human
than do ferret HI data\cite{gupta2006}. When the \textit{p}$_\textrm{epitope}$ between the current vaccine strain and consensus strain of the new cluster is larger than $0.19$, expected vaccine efficacy decreases to 0 for H3N2 influenza, and the current vaccine cannot
be expected to provide protection from new strains. As the examples shown below, our method can detect an incipient dominant strain at its very early stage, and the method appears to require about 10 sequences in the new cluster
for detection.

\subsection*{Demonstration of low-dimensional sequence clustering method. }

\begin{figure}
\begin{center}
\includegraphics[width = 0.43\textwidth]{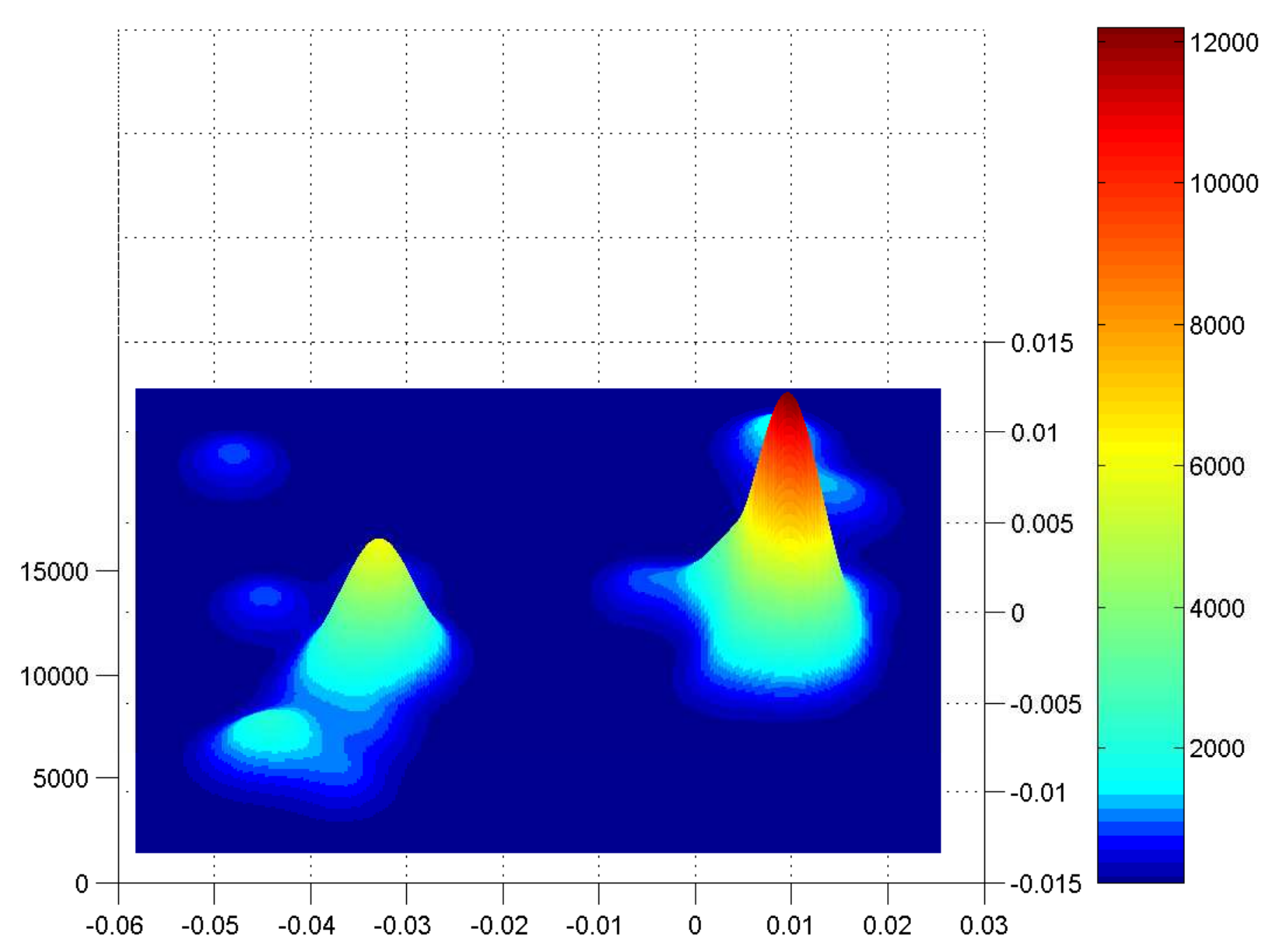}
\includegraphics[width = 0.43\textwidth]{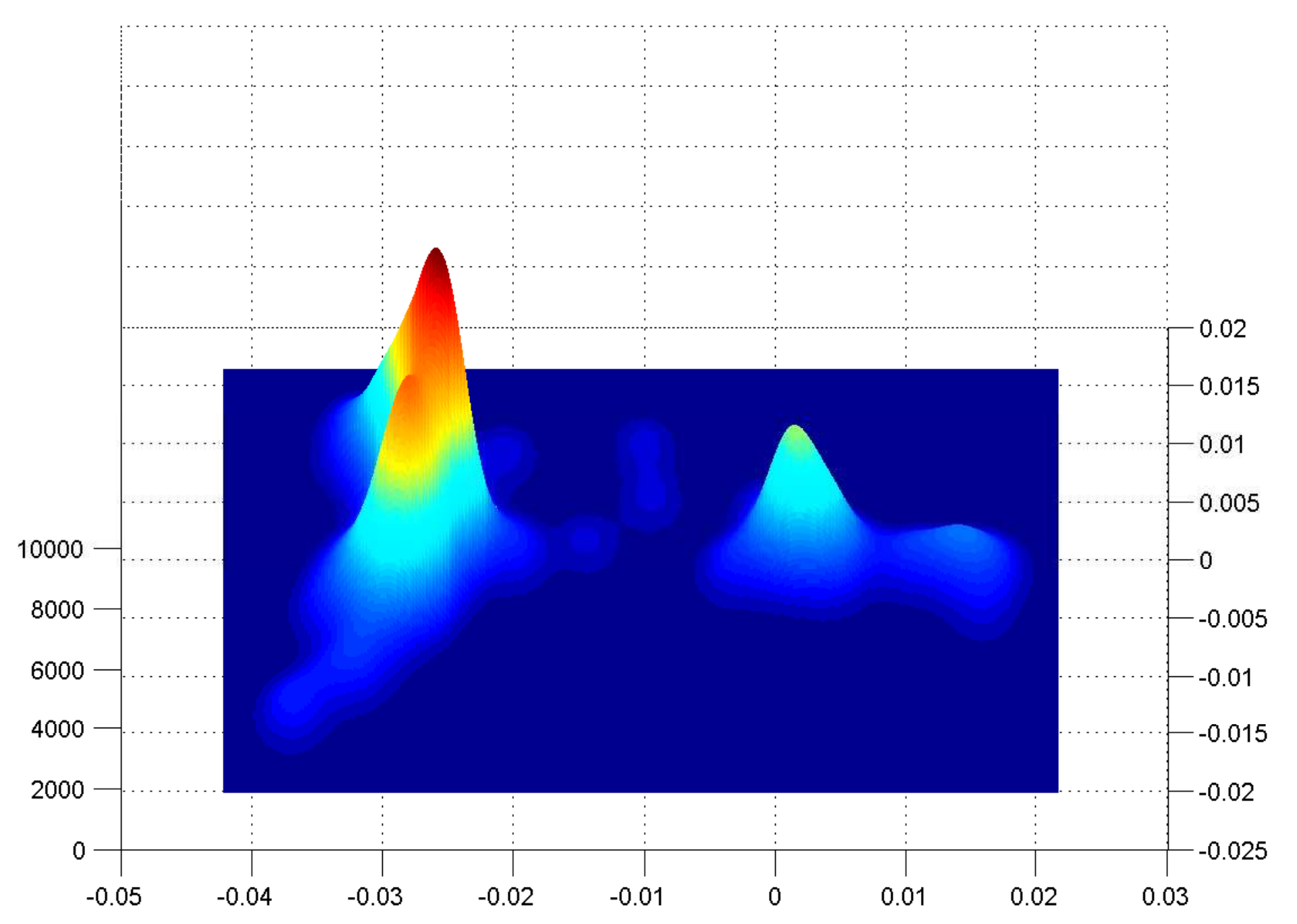}\\
(a)  \hspace{2.4in}  (b)\\
\includegraphics[width = 0.43\textwidth]{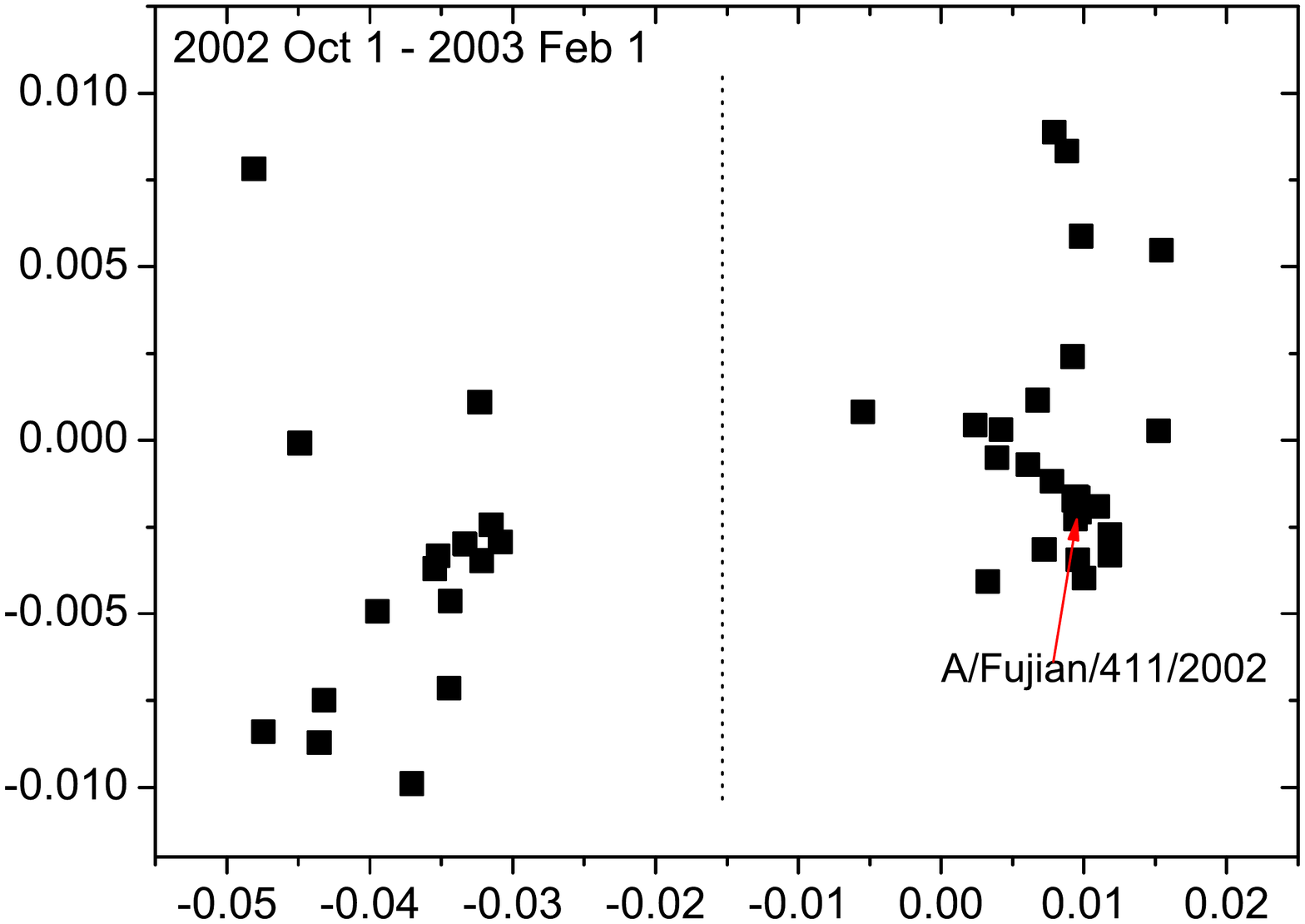}
\includegraphics[width = 0.43\textwidth]{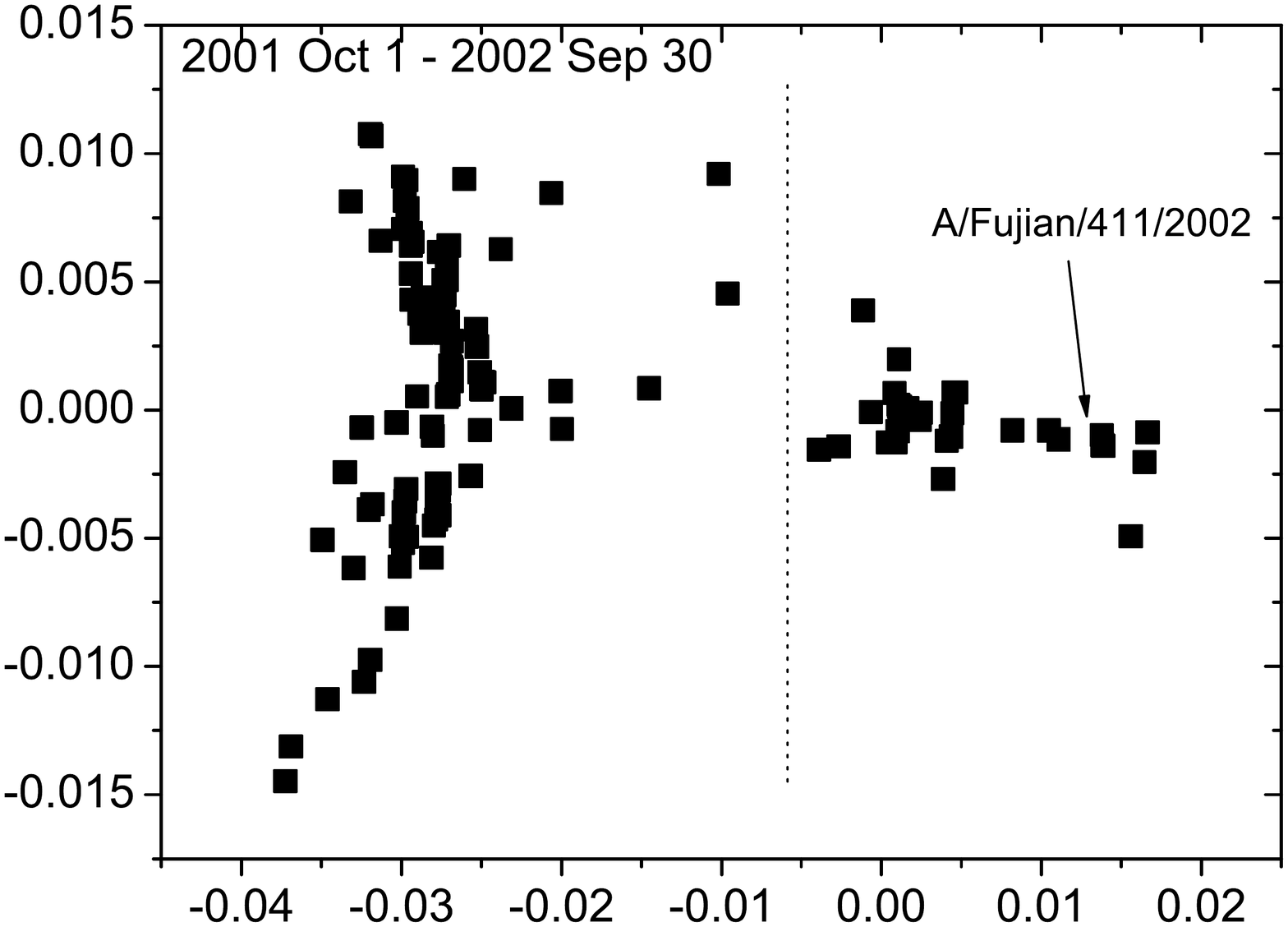}\\
(c)  \hspace{2.4in}  (d)\\
 \caption{(a) Kernel density estimation and (c) protein distance map for H3N2 viruses between October 1, 2002 and
 February 1, 2003. (b) Kernel density estimation and (d) protein distance map for H3N2 viruses between October 1, 2001, and
September 9, 2002.  We plot a dotted line to separate the two clusters.
  The vertical and horizontal axes of all figures represent protein distance as defined in Equation \ref{eq}. A 0.0030 unit of
protein distance equals one substitution of the HA1 protein sequence of H3N2.
  \label{fujian} }
\end{center}
\end{figure}

We demonstrate the method of detecting the A/Fujian/411/2002(H3N2)
strain. The A/Panama/2007/1999 had been the vaccination strain
for four flu seasons between 2000 and 2004 in the Northern hemisphere. The vaccine strain was
replaced by A/Fujian/411/2002(H3N2) in the 2004-2005 flu season, as
described in Table \ref{comparison}. The vaccine strain in the
2003-2004 season was A/Panama/2007/1999, while the
dominant circulating strain became A/Fujian/411/2002(H3N2). This
mismatch resulted in a large decrease in vaccine efficacy in the
2003-2004 flu season\cite{gupta2006}. The vaccine efficacy is estimated to be only 12\%\cite{Dolan2004}.
 We  test whether our method
can detect A/Fujian/411/2002(H3N2) as an incipient dominant strain
before it actually became dominant. We  use only virus sequence data before October 1, 2003. We did not use any virus data collected in 2003-2004 season. Therefore, our prediction and results are made without any knowledge from what  happened in the 2003-2004 season.
 We plot the protein distance map
of the 2001-2002 flu season in Fig.\ \ref{fujian}(d). To detect the
clusters, we plot the kernel density in Fig.\ \ref{fujian}(b) for
the data in Fig.\ \ref{fujian}(d). There are two separate
significant clusters. The  one with the largest kernel density on
the left contains the current dominant strain A/Panama/2007/1999 and
the widespread A/Moscow/10/1999 strain. The smaller one  on the
right is a new cluster, which contains A/Fujian/411/2002. Using the
data as of September 30, 2002, we seek to determine whether the new cluster on
the right in Figs.\ \ref{fujian}(b) and (d) will be the next
dominant strain after A/Panama/2007/1999. We determine whether this
cluster fulfills the two criteria above. First,
this new cluster can be significantly detected by kernel density
estimation. This cluster is separate from the current dominant
strain, as we can see in figure.  Second, we calculated the average
\textit{p}$_\textrm{epitope}$ of the new cluster on the right with
regard to A/Moscow/10/1999, A/Panama/2007/1999 and A/Fujian/411/2002
to be 0.214, 0.1214, and 0.083, respectively. This means the current vaccine contains
A/Moscow/10/1999 is expected to provide little protection against viruses in the new cluster.
 This result makes the new cluster fulfill the second criterion.
Thus, we  predict based on the data as of September 30, 2002, that the
cluster on the right in Fig.\ \ref{fujian}(d) will be the next
dominant cluster. This prediction was made on data collected one year
earlier than when the A/Fujian/411/2002 became dominant in the
2003-2004 season. To further support our prediction, in Fig.\
\ref{fujian}(c), we plot the protein distance map from October 1, 2002, to
February 1, 2003, right before the WHO selected the vaccine strain for 2003-2004
season. To detect the clusters, we plot the kernel in Fig.\
\ref{fujian}(a) for the data in Fig.\ \ref{fujian}(c). There are
two separate major clusters observed in the kernel density
estimation in Fig.\ \ref{fujian}(a). The left cluster has the
current dominant strain of A/Panama/2007/1999 and also
A/Moscow/10/1999. The right cluster has the A/Fujian/411/2002.   We
calculated the average \textit{p}$_\textrm{epitope}$ of the right
new cluster with regards to A/Moscow/10/1999, A/Panama/2007/1999,
and A/Fujian/411/2002 to be 0.2725, 0.1811, and 0.0367 respectively.
This result further supports the prediction that the new cluster will become
dominant, and A/Fujian/411/2002, which is the most frequent strain
in the new cluster, will be or is very close to the next dominant strain.
This suggestion proceeds the vaccine component switch by 1-2 years, as shown in Table \ref{comparison}.

\subsection*{Prediction for H3N2 influenza in 2009--2010. }

\begin{figure}
\begin{center}
\includegraphics[width = 0.43\textwidth]{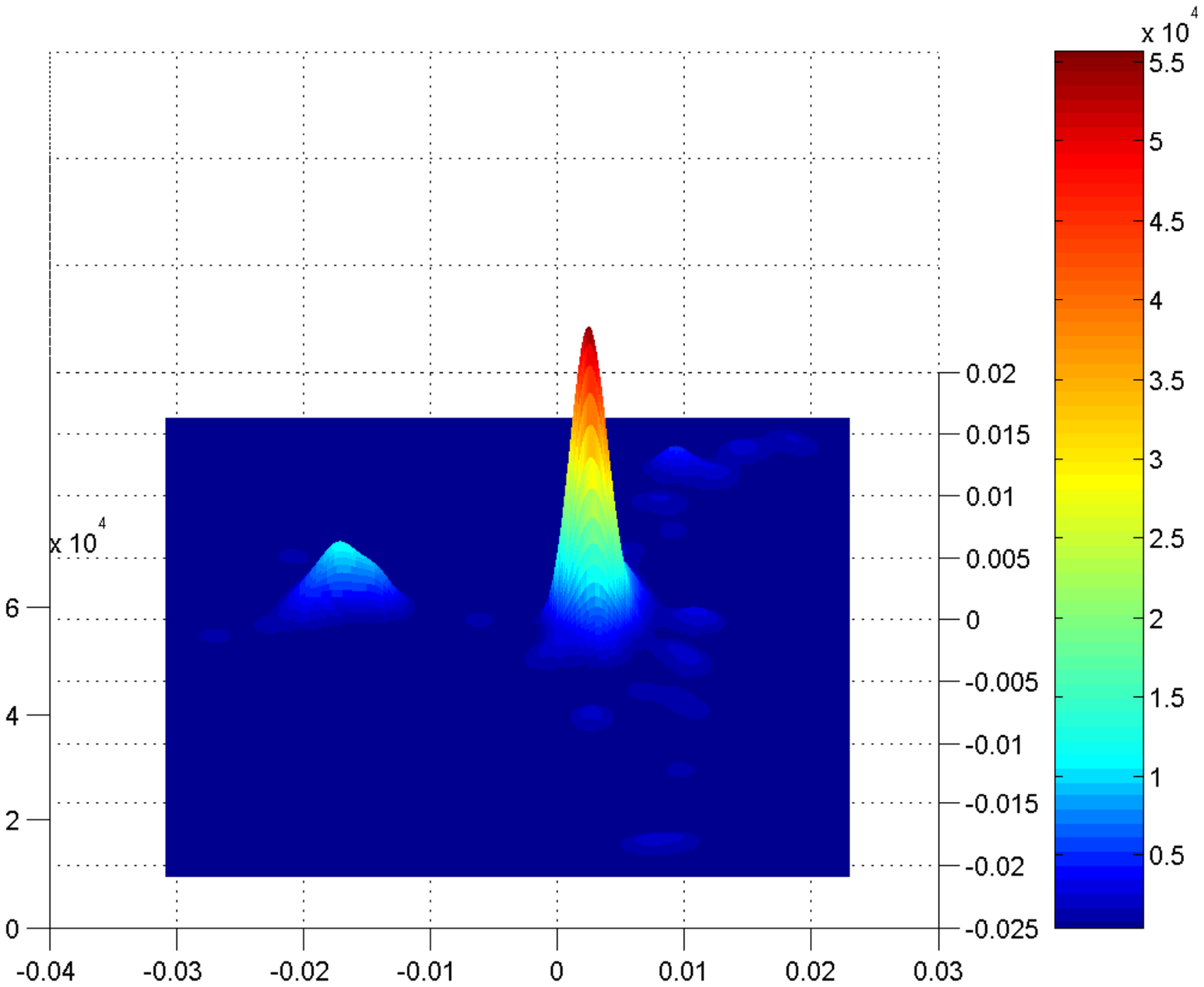}
\includegraphics[width = 0.43\textwidth]{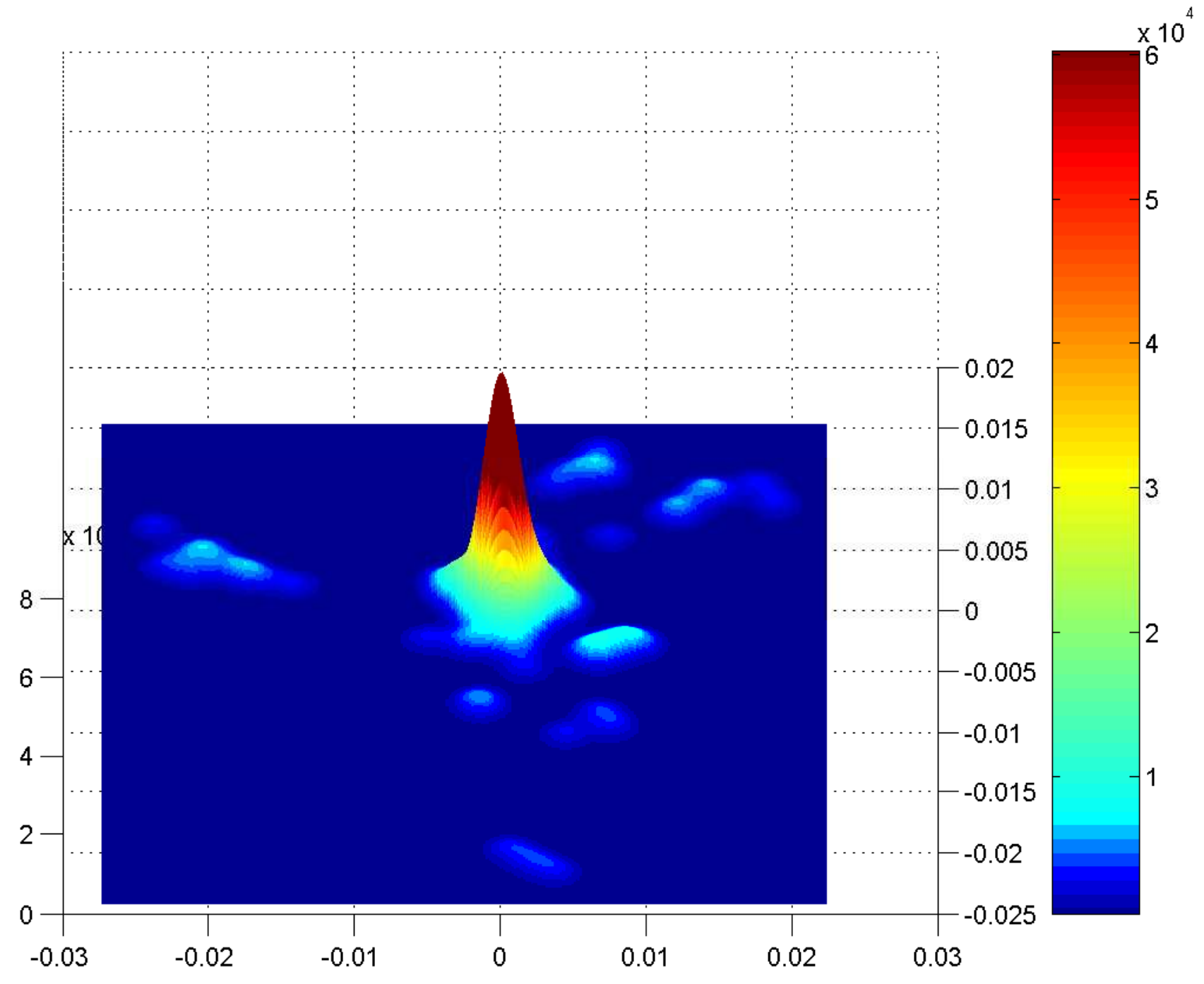}\\
(a)  \hspace{2.4in}  (b)\\
\includegraphics[width = 0.43\textwidth]{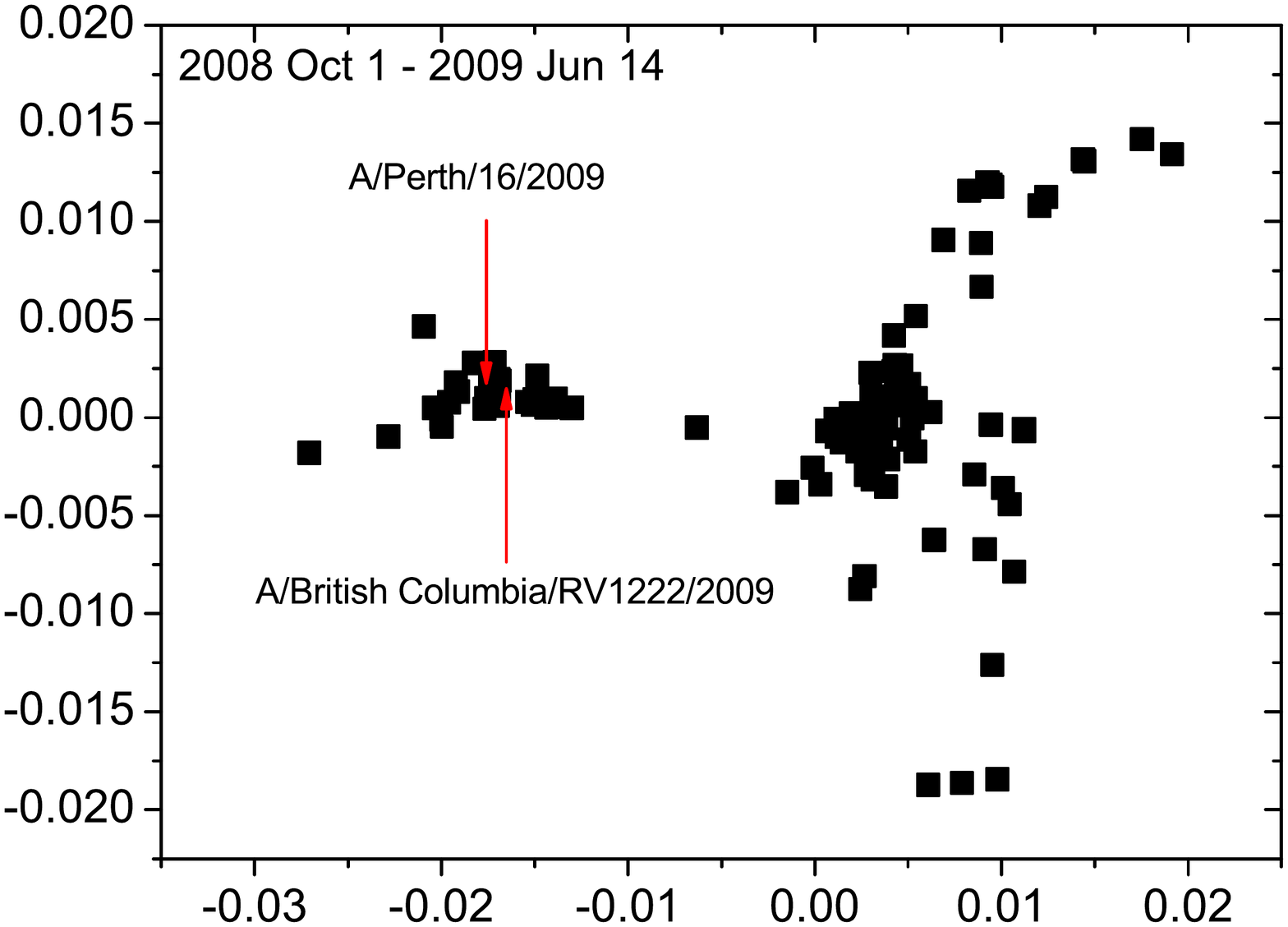}
\includegraphics[width = 0.43\textwidth]{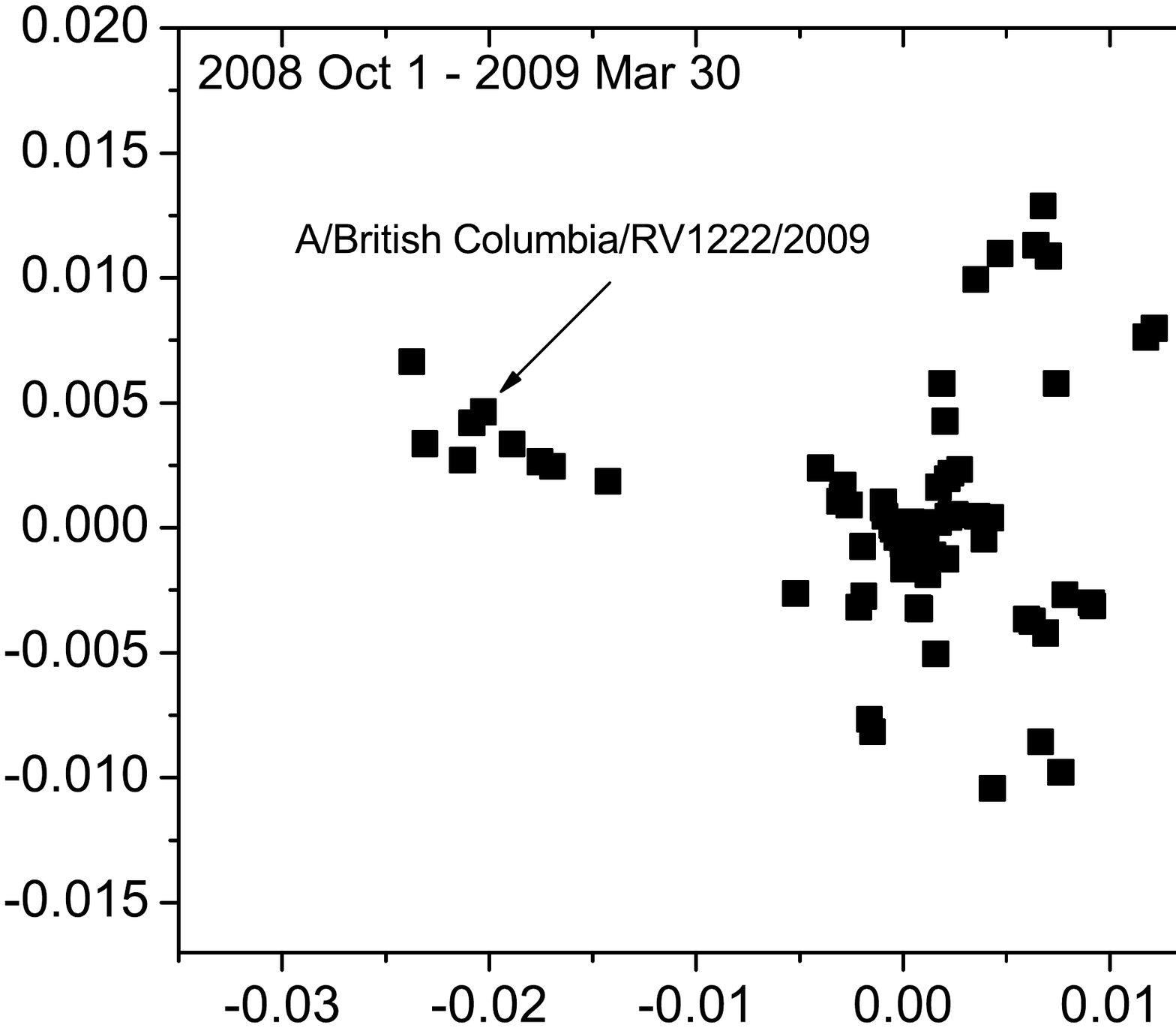}\\
(c)  \hspace{2.4in}  (d)\\
 \caption{(a) Kernel density estimation and (c) protein distance map for H3N2 viruses from
 October 1, 2008, to June 14, 2009. (b) Kernel density estimation and (d) protein distance map for H3N2 viruses between October 1, 2008, and  March 30, 2009. The vertical and horizontal axes of all figures represent protein distance as defined in Equation \ref{eq}. A 0.0030 unit of
protein distance equals one substitution of the HA1 protein sequence of H3N2.
  \label{novel} }
\end{center}
\end{figure}

By applying our method to the 2008-2009 flu season, we predict that
the dominant H3N2 strain in the 2009--2010 flu season may switch.
Based on the flu activity in the 2008-2009 flu season, the WHO made
the recommendation in February 2009 that A/Brisbane/10/2007(H3N2)
should be used as the vaccine\cite{whovaccine}. However, a new strain
evolved just after the recommendation was published. The British
Columbia Center for Disease Control detected a new virus
strain\cite{promed1,promed2} with 3 mutations in antigenic
sites (two in epitope B and one in epitope D).
Since this new strain is relatively far from the vaccine strain,
with \textit{p}$_\textrm{epitope}=0.095$,
vaccine efficacy is expected to decrease to 20$\%$\cite{gupta2006,deem2009}.
However, since the mutations in this new strain ``do not fulfill the criteria proposed
by Cox as corresponding to meaningful antigenic drift"\cite{promed1,cox1995}, and this strain
 still remained the minority of H3N2 viruses  in July 2009, health
authorities were not certain that this new strain would replace the
current dominant strain in 2009--2010 flu season. We use our method
to investigate whether this new strain will be the next dominant
strain. We construct the protein distance map as shown in Fig.\
\ref{novel}(c). We plot the kernel density estimation in Fig.\
\ref{novel}(a) for data in Fig.\ \ref{novel}(c). By the data up to
June 14, 2009, we see two major clusters in Fig.\
\ref{novel}(a). The larger one on the right contains the current
dominant strain A/Brisbane/10/2007, and the left one is a new
cluster which contains A/British Columbia/RV1222/2009. It is
apparent that this new cluster is separate from the current
dominant cluster. Thus, this cluster fulfills the first criterion. We
calculated the average of \textit{p}$_\textrm{epitope}$ of strains
in the left new cluster with regards to A/Brisbane/10/2007 and
A/British Columbia/RV1222/2009 to be 0.103 and 0.042 respectively.
The vaccine that contains  A/Brisbane/10/2007 has an expected efficacy of 20\%
to the virus strains in the new cluster.
Thus, this new cluster satisfies both two criteria, and so we predict
that this cluster which contains A/British Columbia/RV1222/2009 will
be the dominant cluster in the 2009--2010 season. The earliest time for us to make this
prediction is March 30, 2009. In Fig.\ \ref{novel}(d) and
(b), we already see this new cluster on the left side of figure,
though since there are only about 10 sequences in the new cluster,
the kernel density of this new cluster is  smaller than that in the
dominant cluster. This strain was mentioned as a
concern on 5 May 2009, although by conventional methods the strain
was not considered a potentially new dominant strain in July 2009\cite{promed1}.
With the method of the present paper, this new cluster is suggested earlier using
the data as of March 30, 2009.

\subsection*{Comparison with previous results.}

Here we present a historical test of the method. For each flu season in the
North Hemisphere from 1996,
we use only the H3N2 sequences data until February 1, before WHO published the recommendation for vaccine. We use
the low dimensional clustering to made the prediction for the dominant strain. The conventional method as used by WHO is phylogenetic analysis combined with ferret antisera HI assay. In Table \ref{comparison}, we compare the method with the conventional method. In the most recent 14 flu seasons,
influenza subtype H3 was dominant in 10. The WHO H3N2 vaccine component matches the circulating strains in 8 seasons. Our predictions match the circulating strains in 9 seasons.
In 1997-1998 season, a novel flu strain Sydney/5/97 was found in June 1997. Because no similar strains were collected  before February 1, neither of the two methods can predict it.
In 2003-2004 season,
our method predicts Fujian/441/2002 as the dominant strain,
while phylogenetic analysis combined with ferret antisera HI assay did not.
 For all other 8 seasons dominated by  influenza subtype H3, the predictions of both methods matched the dominant
circulating strain. The 2009--2010
influenza season was dominated by H1N1. But data from local outbreaks of H3N2 infections\cite{promed1,promed2} showed that the dominant H3N2 strain was A/British Columbia/RV1222/2009, as  predicted in Table \ref{comparison},
 rather than the vaccine strain A/Brisbane/10/2007.
For the 2010-2011 season, we recommend A/British Columbia/RV1222/2009 as a vaccine strain, and the WHO recommended A/Perth/16/2009. These two strains are in the same cluster and antigenically similar with a small \textit{p}$_\textrm{epitope}=0.048$. Although these two strains are slightly different, the vaccine is expected to be effective.

\subsection*{Detecting A/Wellington/1/2004 in the
2004 flu season in the Southern hemisphere}
The low-dimensional clustering can also be applied to influenza in the
Southern hemisphere. As an example, we test our method on the
2004 flu season. The recommended H3N2 vaccine strain by WHO used in the
2004 flu season in the Southern hemisphere was A/Fujian/411/2002. Data from
the surveillance network suggested that the circulating dominant flu strain in
the 2004 season in Southern hemisphere was A/Fujian/411/2002, and a late
surge of A/Wellington/1/2004 was also observed. For example, in Argentina,
a study showed that about 50\% of infections were closely related to
A/Fujian/411/2002 and another 50\% were closely related to
A/Wellington/1/2004\cite{sant2008}. In New Zealand, the dominant flu strain
was A/Fujian/411/2002 which caused 78\% of flu infections\cite{newzealand2},
and a late season surge of A/Wellington/1/2004 was also reported\cite{promed3}.
Therefore, the vaccine recommended by WHO matches the dominant strain and
would be expected to have vaccine efficacy in the
2004 season in Southern hemisphere.

We here use the low-dimensional clustering method to detect the
A/Wellington/1/2004 strain, which is not the major dominant strain but
caused significant infections in the 2004 flu season.
 We plot the protein distance and kernel density estimation for the
H3N2 viruses in Fig.\  6(d) and 6(b). We use the data only
as of February 1, 2004, 3 months prior to the 2004 flu Southern hemisphere
season, which is
usually from May to September.
We observed two clusters. The major cluster on the left side of
Fig.\ 6(d) is A/Fujian/411/2002-like, which was the
vaccine strain in 2004 season. There is a new cluster in the
right side of Fig.\ 6(d) which contains A/Wellington/1/2004.
  The \textit{p}$_\textrm{epitope}$ of A/Wellington/1/2004 with regards
to A/Fujian/411/2002 is 0.118. Therefore, we predict that
A/Wellington/1/2004 will infect a large fraction of the population, and the
 A/Fujian/411/2002 vaccine is expected to provide only partial
protection against the A/Wellington/1/2004 virus.
However, since the appearance of A/Wellington/1/2004 was
just before the the 2004 flu season, it did not have sufficient time to
spread out and become the dominant strain in the 2004 flu season. From our
observation, it usually takes about 8 months or longer for a new strain to
become dominant after its appearance in a new cluster. Therefore, the
predominant flu strain in 2004 season is expected to
be A/Fujian/411/2002 based on the data as of February 1, 2004. This result
agrees with the dominant flu strain  in the 2004 flu season.

\subsection*{Detecting A/California/4/2004 as a future dominant strain}
As a further example of
applying the low-dimensional clustering method to
influenza in Southern hemisphere,
we test the method on the 2005 flu season.
The recommended H3N2 vaccine strain in the 2005 flu season
in the Southern hemisphere was
A/Wellington/1/2004. Data from HI assay tests and surveillance suggest
that the dominant H3N2 strain in the 2005 season was A/California/7/2004.
In HI tests with postinfection ferret sera the majority of
influenza A(H3N2) viruses from February 2005 to October 2005 were
closely related to A/California/7/2004, as reported by WHO on
October 7, 2005\cite{wer05s}. Surveillance data from Victoria, Australia,
show that 45\% of influenza A infections were A/California/7/2004-like (H3),
11\%  were A/Wellington/1/2004 (H3) and 44\%  were A/New Caledonia /20/99-like
(H1), as collected in the 2005 flu season\cite{turner2006}. Surveillance data
from New Zealand also show that the dominant H3N2 strain in the 2005 flu season
was A/California/7/2004\cite{newzealand}.

We plot the protein distance for
the H3N2 viruses in the 2003-2004 flu season in Fig.\
6(c). We only use the data as of September 30, 2004, earlier
 than the October 2004 date when the WHO published
the influenza vaccine recommendation for Southern hemisphere.
 We plot the kernel density estimation in Fig.\
6(a) for the data in Fig.\ 6(c).
There are three major clusters in Fig.\ 6(a). The
one on the left is the current dominant cluster which are mostly A/Fujian/422/2002-like viruses.
There is a middle cluster centered on A/Wellington/1/2004.
The one on the right contains A/California/7/2004.
Both the A/California/7/2004 cluster and the
A/Wellington/1/2004 cluster are antigenically novel from A/Fujian/411/2002.

When the protein distance map  and kernel estimation as of February 1, 2004, is plotted in
Fig.\ 6(d) and (b), we still see the A/Wellington/1/2004 cluster. With these data, the A/California/7/2004 cluster is no longer observed. Thus, A/California/7/2004 cluster is a newly appearing cluster and
we consider it to be the emerging strain.
The new cluster which contains A/California/7/2004 is
separate from the current dominant cluster.
We calculated the
average \textit{p}$_\textrm{epitope}$ of the new cluster that contains
A/California/7/2004
with regard to A/Fujian/411/2002 to be 0.112.
This makes the new
cluster fulfill both criteria for an
incipient dominant strain cluster. So we predict based on the
information as of September 30, 2004, that A/California/7/2004 will be the
next dominant strain after A/Fujian/411/2002 in Southern hemisphere.
We further predict from these data that A/California/7/2004 will be the
dominant strain in the following flu season in the Northern hemisphere.
These predictions agree with the observed dominant strain in the 2005 flu season.

\section*{Discussion}
The evolution of influenza virus is driven by cell receptor distributions, non-specific innate host defense
mechanisms, cross immunity\cite{gupta1998,ferguson2003}, and other contributions to viral fitness. In this paper, we focused on HA protein evolution under antibody selection pressure.  The degree to which the immunity induced by one strain protects against another strain depends on their antigenic distance\cite{gupta2006}. Because the human
immune response to viral infection is not completely cross protective,
natural selection favors amino-acid variants of the HA protein that allow the virus to evade
immunity, infect more hosts, and proliferate. Mutant strains surround the dominant strain and group into a cluster rather than evolve in a defined direction\cite{Smith2004,Joshua2002}.  After the virus has circulated in population for one or more years, effective vaccines and cross immunity of the population drive the evolution of influenza by mutation and reassortment. This evolution increases the immune-escape component of the fitness of new strains, and eventually causes a new epidemic. These new immune-escape strains will form a new cluster, and the old clusters will die out, thus starting a new cycle. This process of  creating of new clusters is what our method detects.

The low dimensional clustering can  be used not only in genetic sequences but also on distances calculated from inhibition assays of antibody and antigens, as developed by
 Lapedes and Farber\cite{lapedes2001} and
Smith \emph{et al.}\cite{Smith2004}. The inhibition assay provides an approximation of antigenic distance and is broadly used as a marker for vaccine efficacy. The inhibition assay suffers from low resolution of data, which multidimensional scaling improves, and is less able to predict the vaccine efficacy than the \textit{p}$_\textrm{epitope}$  method\cite{gupta2006}. The genetic sequences used here are a direct description of the evolution of pathogen and antigenic distance of influenza. To aid vaccine selection, the low dimensional clustering on genetic sequences appears informative.

Challenges may arise in application of the method described here. If two or more new clusters appear in one season, additional information is needed to decide which cluster should be chosen for vaccine. Fortunately, it has been shown that the evolution of influenza is typically in one direction\cite{ferguson2003,Smith2004}. It is rare to have two or more new clusters in the protein distance map in one season. As experience with the low dimensional sequence clustering is gained, it may be that cluster structure will allow more precise prediction of vaccine efficacy. Despite these issues, the method described here can assist the design of vaccines, and it provides a new tool to analyze influenza viral dynamics.
We did not see any false positive results in Table \ref{comparison}.

The current WHO method works quite well in many years. The
method discussed here appears to offer an additional tool
which may provide additional utility.

\section*{Materials and Methods}

\subsection*{Data sources}
Influenza hemagglutinin A(H3N2) sequences before October 1, 2008, and
A(H1N1) sequences as of December 5, 2009, were downloaded from NCBI
Influenza Virus Resources\cite{bao2008}. All hemagglutinin sequences
used in our study are filtered by removing  identical sequences,
Thus, all groups of identical sequences in the dataset are
be represented by the oldest sequence in each group. This approach
 reduces the number of sequences by keeping only the unique
sequences in the dataset. The hemagglutinin proteins of 2009 A(H1N1)
used in our work are listed in Supplementary Table 3.  The numerical labels
in Figs.\ 1 and 2 are the same as the labels in the first column of
Supplementary Table 3. Influenza A(H3N2) sequences after October 1, 2008,
were downloaded from GISAID database, see Supplementary Table 6. GISAID has the latest H3N2 sequence data.

\subsection*{Geographical spread pattern of 2009 A(H1N1)}
It is believe that the 2009 A(H1N1) virus was most likely originated from
Mexico\cite{Fraser2009}. It first spread to the neighboring country
USA and then to other countries. We display this geographical
spread pattern in Fig.\ 1. We take the founder-F1
relationship from Fig.\ 1, and assume the virus spreads from
location of founder to the location of F1. We consider three regions: USA, Mexico and other countries except USA and Mexico. Then we count  the cases of spreading
from one region to another region.  In Supplementary Table 2, we show that
we observed many more paths of spreading from the USA to other
countries than  from other countries to the USA. The major  path of
spreading is from USA to other countries. This result indicates our
directional evolutionary map of Fig.\ 1 is in good agreement
with the pattern of geographical spread.

\subsection*{Multidimensional scaling}
The goal of multidimensional scaling is to represent the distance
of proteins by a Euclidean distance in coordinate space. We
calculate the distance
between proteins $i$ and $j$, $d_{ij}$, by the number of amino acid residue
differences divided by the total number of amino acid residues, as defined by Equation 1 in the main text.
To do multidimensional scaling, we start with the distance of the proteins.
 The object of multidimensional scaling is to find the two, or $p$ in general, directions
 that best preserve the distances $d_{ij}$ between the $N$ proteins
\begin{equation}
F=\sum_{i,j=1}^{N}(d_{ij}-D_{ij})^2
\end{equation}
 Here, $D_{ij}=\|x_i-x_j \|$ is the  Euclidean distance between proteins $i$ and $j$ in the projected space, and $\|\bullet\|$
is the vector norm.
The algorithm is as follows. Let the matrix $A=[(a_{ij})]$, where $a_{ij}=-\frac{1}{2}d_{ij}^2$. The eigenvalues of A are $\gamma_1, \gamma_2,..., \gamma_N$ and  $\gamma_1 \geq \gamma_2 \geq...\geq\gamma_N$. Let $V^{(1)}=(v_1^{(1)}, v_2^{(1)},...,v_N^{(1)})$ be the eigenvector of $\gamma_1$ and $V^{(2)}=(v_1^{(2)}, v_2^{(2)},...,v_N^{(2)})$ be the eigenvector of $\gamma_2$. Let $x=\sqrt{\gamma_1} V^{(1)}$ and $y=\sqrt{\gamma_2} V^{(2)}$. The two coordinates in Figs.\ 2--6 are $x$ and $y$. The $x$-axis in the protein distance map is the largest eigenvector. We take H3N2 2008--2009 season as an example. In Fig.\ 5(c),
we observe  two clusters. One cluster is on the right side of figures with $x$ value positive and another one has negative $x$ values.  We define the consensus sequence of a group of flu strains by taking the most frequent amino acid at each position. We calculate the consensus sequences both for the strains in the cluster on the right and on the left of figure. We found amino acids at four positions (76, 160, 172, 203) are different for these two consensus H3N2 strains, see Supplementary Table 1. Interestingly, the Shannon entropy calculated from all 2008--2009 season sequences at these four positions (0.43, 0.67, 0.59, 0.50) are the largest, which means the diversity at these four position are the largest.

There is software available to run the multidimensional scaling.
We use the Matlab function ``CMDSCALE" to generate an $N\times p$ configuration matrix $Y$.
Rows of Y are the coordinates of N points in $p$-dimensional space.
The ``CMDSCALE" also returns a vector $E$ containing the sorted eigenvalues of
what is often referred to as the ``scalar product matrix," which, in the simplest case, is equal to YY$^\top$.
If only two or three of the largest eigenvalues $E$ are much larger than others,
then the  matrix $D$ based on the corresponding columns of Y nearly reproduces the original distance matrix $d$.
We used the influenza H3N2 in 2001--2002 season as an example. The five largest of all 180 eigenvalues  are 0.0361,
0.0032, 0.0024, 0.0020, 0.0016. The first two largest eigenvalues contribute
70\% to the sum of all 180 eigenvalues, which indicates $p=2$.
Then, we plot the the $N$ points in a two-dimensional graph. Each
point represents a protein. The Euclidean distance between any two points
$D_{ij}$ on the graph should be equal to or close to the distance
of these two proteins. that is,  $D_{ij}\approx d_{ij}$. As an example, in Supplemental Fig.\ 1,
We show that $D_{ij}$ and $d_{ij}$ have a strong linear relationship.
A short MATLAB program of multidimensional scaling is as follow.
\\
\texttt{
\% Multidimensional scaling. \\
\% alignment.aln is a sequence multialignment file \\
\% generated by software ClustalW.\\
clear\\
Sequences = multialignread('alignment.aln');\\
distances $=$ seqpdist(Sequences,'Method','p-distance');\\
Y $=$ cmdscale(distances);\\
scatter(Y(:,1), Y(:,2));}

\subsection*{\textit{p}$_\textrm{epitope}$ estimation}
The value of \textit{p}$_\textrm{epitope}$ is a measure of antigenic
distance between influenza A vaccine and circulating strains.
 The hemagglutinin protein has five epitopes.
The dominant epitope for a particular circulating strain in a particular season was taken as that which
had the largest fractional change in amino acid sequence relative to the vaccine strain.
The value of \textit{p}$_\textrm{epitope}$ is defined as the fraction of number of amino acid differences in the dominant epitope to total number of amino acids in the dominant epitope. The antigenic distance between the vaccine
strain and the circulating strain is quantified by \textit{p}$_\textrm{epitope}$. By a metaanalysis of historical vaccine efficacy data from over 50 publications, Gupta \emph{et al.} showed in a metaanalysis that the \textit{p}$_\textrm{epitope}$ between vaccine strain and circulating strain correlates well with the vaccine efficacy, with $R^2>0.8$\cite{gupta2006}.
The value of \textit{p}$_\textrm{epitope}$ can be easily calculated from sequence data.

\subsection*{Biases in the data}
There are two biases in the sequence data.
First, more isolates are sequenced in recent years. Generally speaking, more
sequences  make the vaccine selection based on low-dimensional
clustering methods more reliable. That is why we compared
low-dimensional clustering methods with WHO results only
since 1996 in Table 1.
To avoid these biases in the generation of
the figure of evolution history of influenza for the 40 years
(Fig.\ 3), we choose 20 random isolates for
each season, even though the database contains more sequences
in recent years.  Second, most isolates are
collected in USA. We
found that many isolates collected in USA are identical, because of the high
sampling rate in USA.
To reduce this bias, we collapse redundant strains,
keeping only distinct strains.

\section*{Funding}
 This research was supported by DARPA under the FunBio program.

\bibliography{jiankui}
\bibliographystyle{plos2009}

\clearpage

\clearpage

 \clearpage

 \clearpage

\begin{figure}
\begin{center}
\includegraphics[width = 0.43\textwidth]{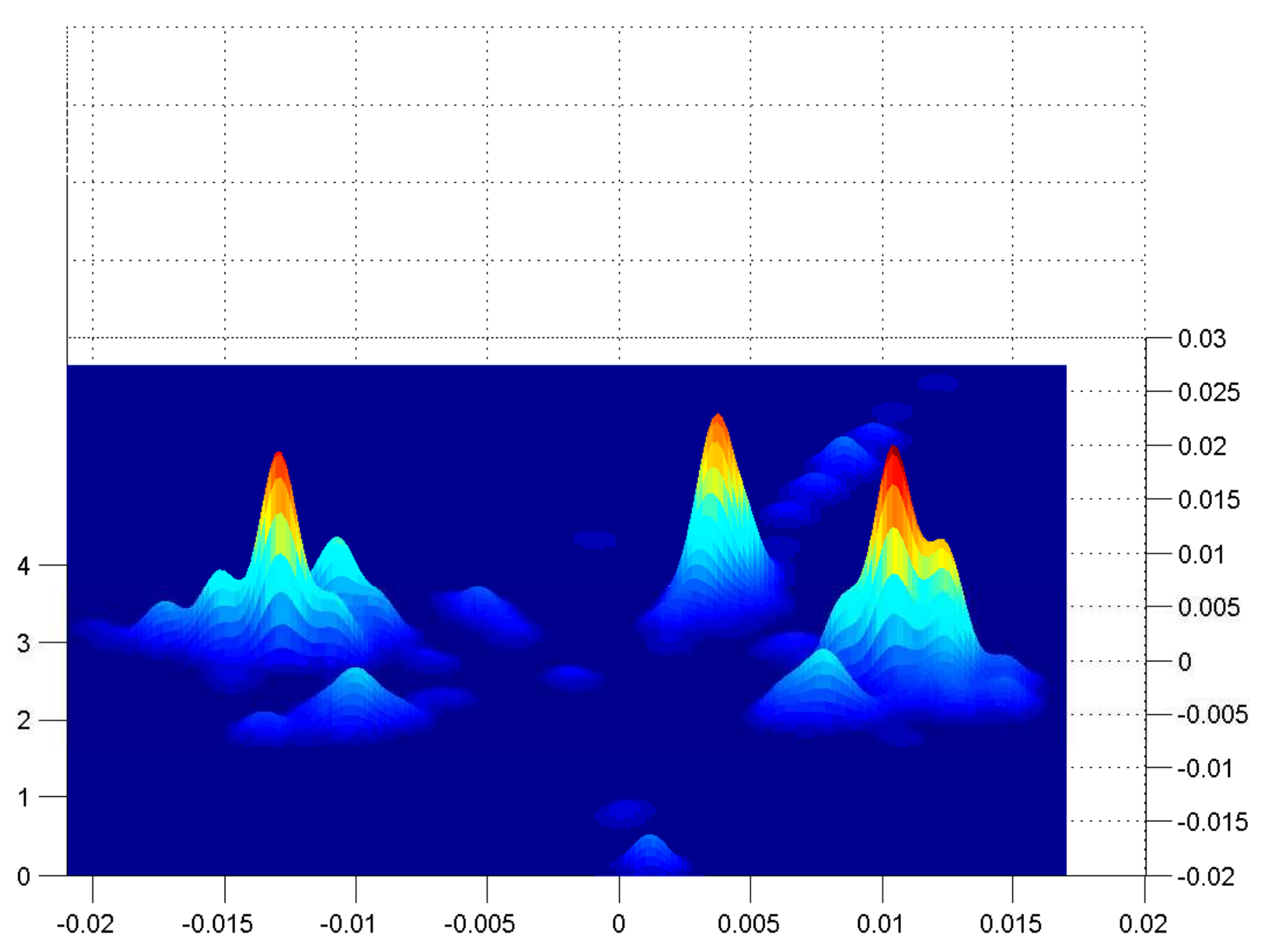}
\includegraphics[width = 0.43\textwidth]{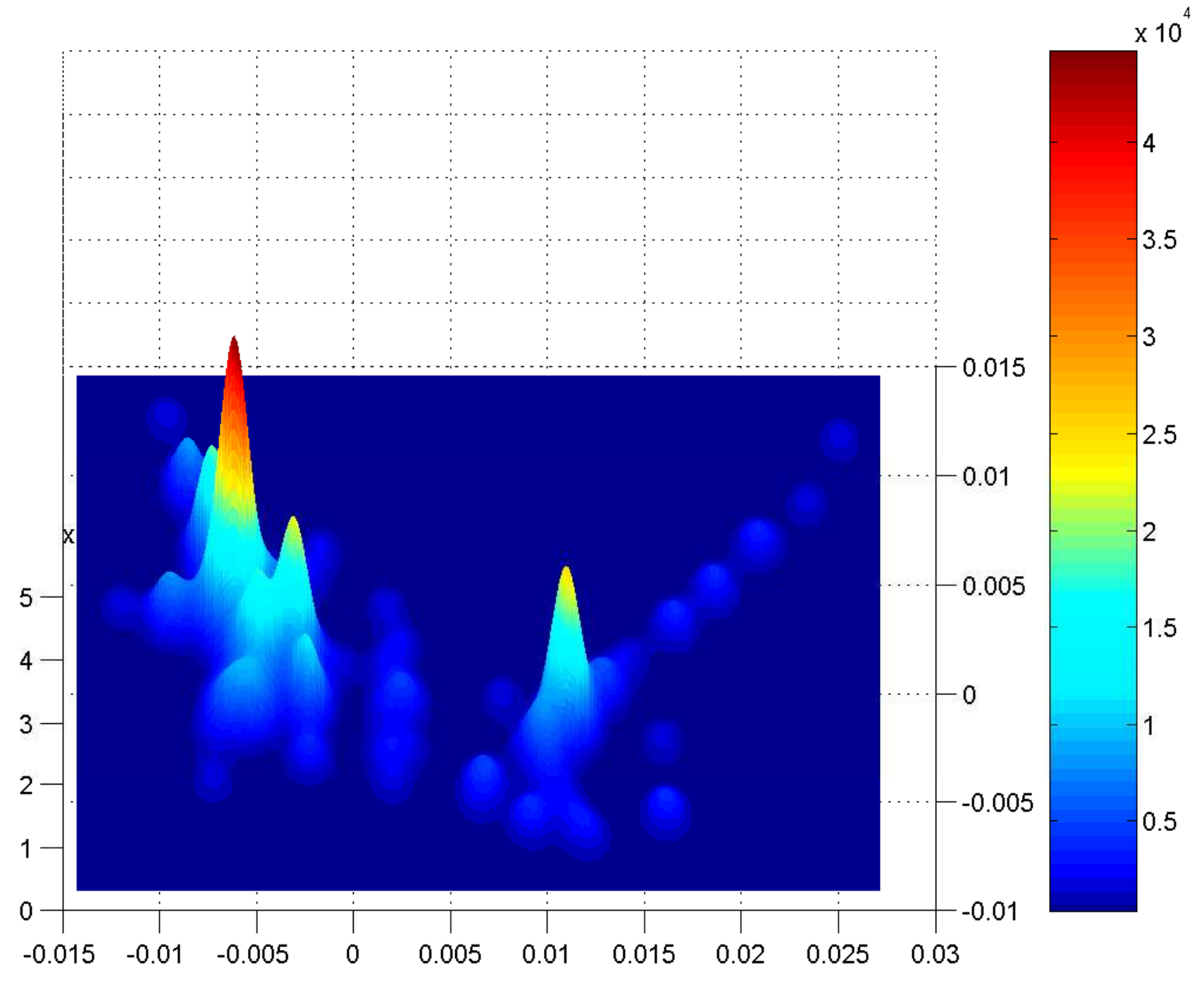}\\
(a)  \hspace{2.4in}  (b)\\
\includegraphics[width = 0.43\textwidth]{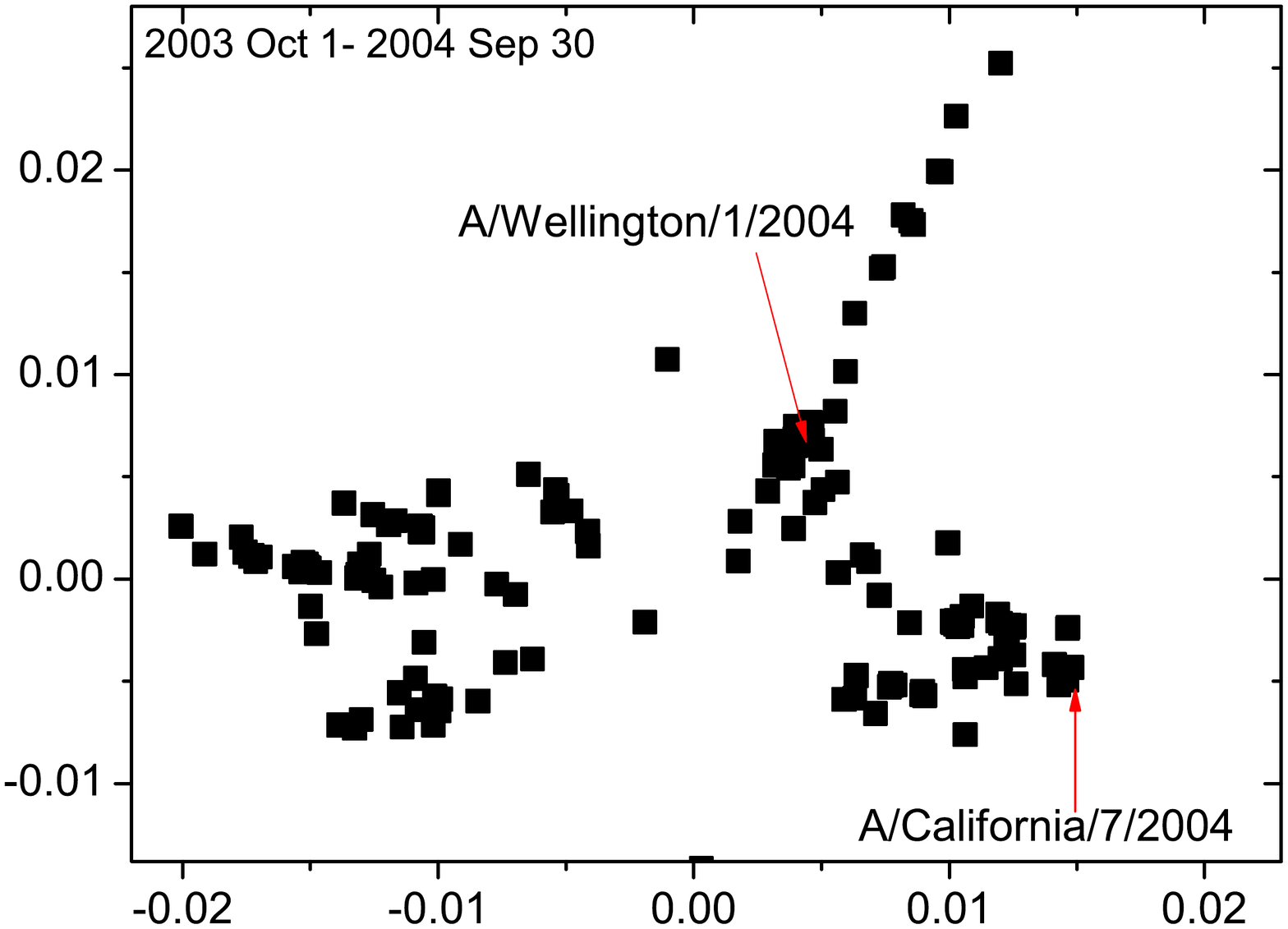}
\includegraphics[width = 0.43\textwidth]{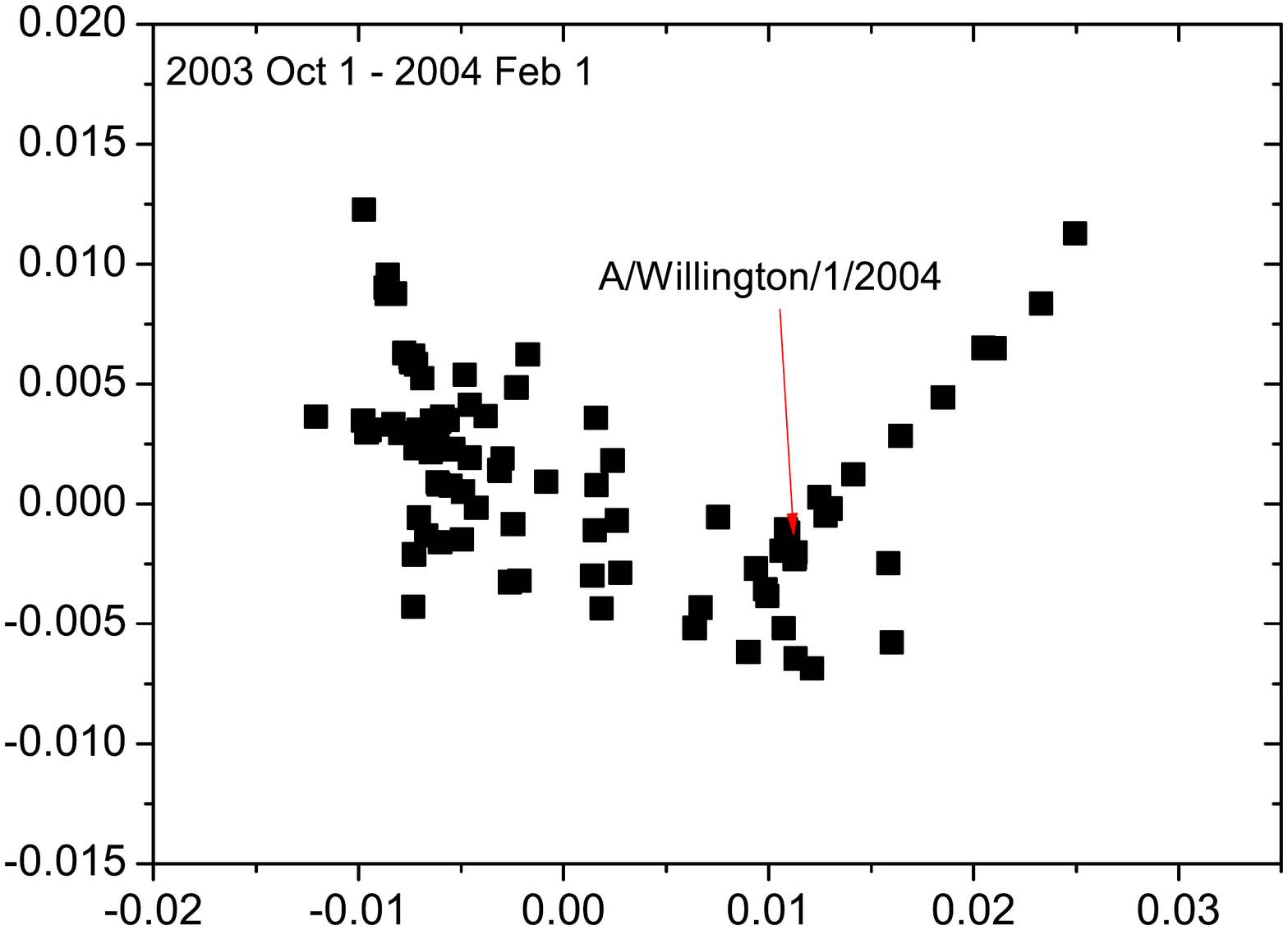}\\
(c)  \hspace{2.4in}  (d)\\
 \caption{(a), Kernel density estimation for the protein distance map for H3N2
viruses between 10/01/2003 and 09/30/2004. (b), Kernel density
estimation for the protein distance map for H3N2 viruses between
10/01/2003 and 02/01/2004. (c), Protein distance map for H3N2
viruses between 10/01/2003 and  09/30/2004. We plot a dotted line to
separate the two clusters. (d), Protein distance map for H3N2
viruses between 10/01/2003 and 02/01/2004. The vertical and horizontal axes of all figures represent protein distance. A 0.0030 unit of
protein distance equals one mutation of the HA1 protein sequence of H3N2.
  \label{california} }
\end{center}
\end{figure}
\clearpage

\clearpage

\begin{table}
\begin{center}
\caption{Summary of results \label{comparison}}
\begin{tabular}{  l  l  l  l  l }
 \hline
 Flu season & Vaccine strain & Our prediction &  Circulating  & Circulating \\
         & from WHO\cite{whovaccine} &	 &	H3N2 strain & subtype\\ \hline
1996-1997 & Wuhan/359/95 &	Wuhan/359/95 &	Wuhan/359/95 & H3\\
1997-1998 &	Wuhan/359/95 &	Wuhan/359/95 & 	Sydney/5/97  & H3\\
1998-1999 &	Sydney/5/97 &	Sydney/5/97 &	Sydney/5/97 & H3\\
1999-2000 &	Sydney/5/97 &	Sydney/5/97 &	Sydney/5/97  & H3\\
2000-2001 &	Panama/2007/1999 & 	Panama/2007/1999 & N/A & H1 \\
2001-2002 &	Panama/2007/1999 & 	Panama/2007/1999 & 	Panama/2007/1999 & H3 \\
2002-2003 &	Panama/2007/1999 &	Fujian/411/2002 & N/A & H1 \\
2003-2004 &	Panama/2007/1999 &	Fujian/411/2002 & Fujian/411/2002 & H3 \\
2004-2005 &	Fujian/411/2002 & Fujian/411/2002 & Fujian/411/2002 & H3 \\
2005-2006 &	California/7/2004 & California/7/2004 & California/7/2004 & H3 \\
2006-2007 &	Wisconsin/67/2005 &	Wisconsin/67/2005 &	Wisconsin/67/2005 & H3 \\
2007-2008 &	Wisconsin/67/2005 &	Wisconsin/67/2005 &	N/A & H1 \\
2008-2009 &	Brisbane/10/2007 & Brisbane/10/2007 & Brisbane/10/2007 & H3 \\
2009-2010 & Brisbane/10/2007 & BritishColumbia/RV1222/09 & BritishColumbia/RV1222/09 & H1\\
2010-2011 & Perth/16/2009 & BritishColumbia/RV1222/09 & N/A & N/A \\\hline
 \end{tabular}
 \end{center}
 This table includes the H3N2 vaccine strains, our prediction of dominant strains, the reported dominant circulating H3N2 strains\cite{wer95,wer96,wer97,wer98,wer99,wer00,wer01,wer02,wer03,wer04,wer05,wer06,wer07,wer08,wer09,wer10}, and the circulating subtypes in the northern hemisphere\cite{wer95,wer96,wer97,wer98,wer99,wer00,wer01,wer02,wer03,wer04,wer05,wer06,wer07,wer08,wer09,wer10}. Circulating H3N2 strains are absent if the dominant subtype is H1 or influenza B. The reported dominant H3N2 strains and circulating subtypes data are from WHO Weekly Epidemiological Record (http://www.who.int/wer/en/).
\end{table}


\end{document}


\begin{flushleft}
{\Large
\textbf{Low-dimensional clustering detects incipient
dominant influenza strain clusters}
}
\\
Jiankui He$^{1}$,
Michael W. Deem$^{1,2}$
\\
\bf{1} Department of Physics \& Astronomy, Rice University, Houston, Texas, USA
\\
\bf{2} Department of Bioengineering, Rice University, Houston, Texas, USA
\\
$\ast$ E-mail: Corresponding mwdeem@rice.edu
\end{flushleft}

\section*{Supplementary Data}
\clearpage

\begin{figure}
\begin{center}
\includegraphics[height=2in]{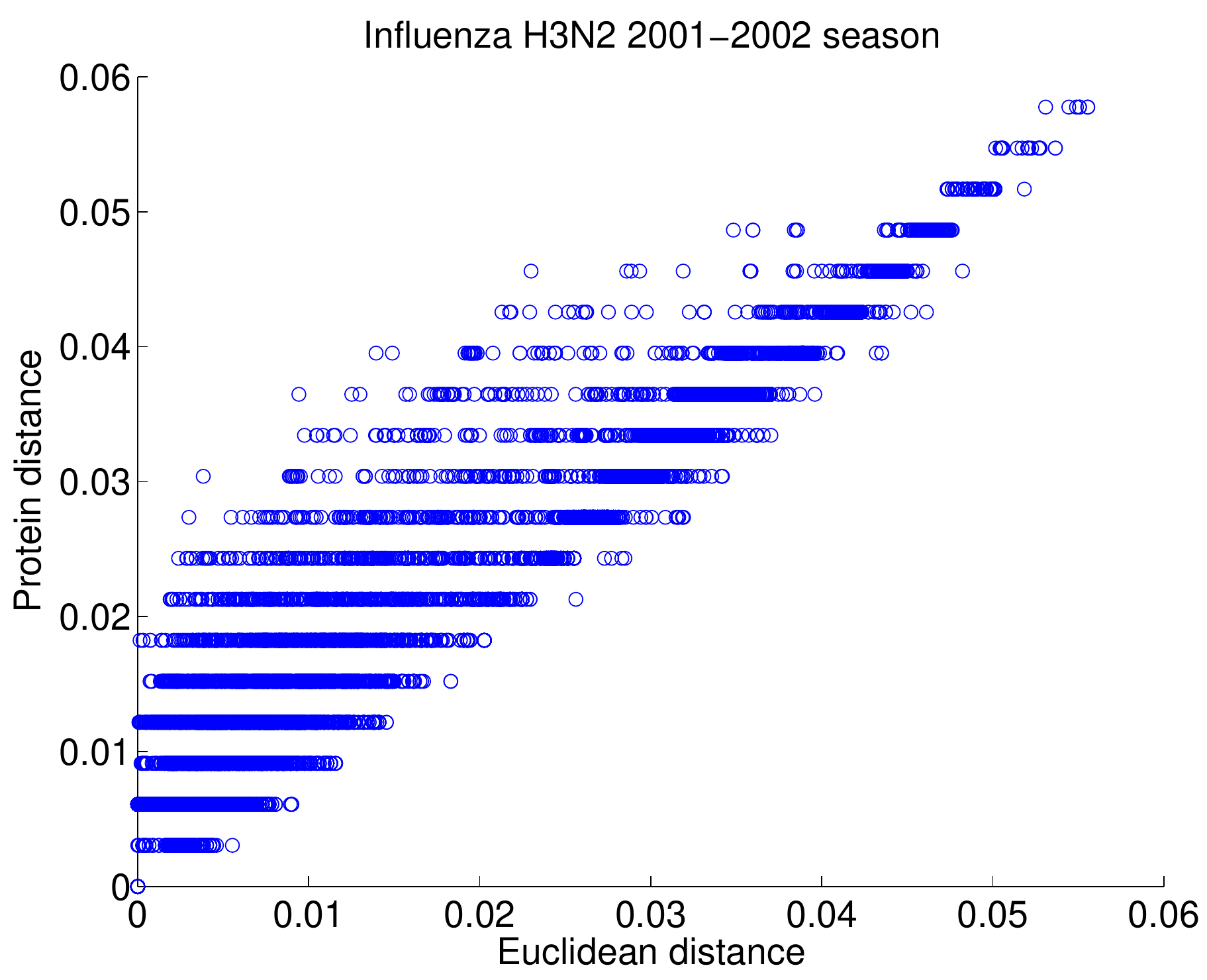}\\
\caption{Plot of Euclidean distances of proteins as in Figure 4(d) of main text on $x$-axis
 and plot of distance of corresponding proteins in $y$-axis. Closeness to the diagonal measures
 fidelity of the low dimensional projection. A 0.0030 unit of
protein distance equals one mutation of the HA1 protein sequence of H3N2.
  \label{eucliean} }
\end{center}
\end{figure}

\clearpage

\begin{table}
\begin{center}
\begin{tabular}{ | l | c |}
 \hline
Spreading path & Number of cases\\ \hline USA to Others & 32
\\\hline Others to USA & 1 \\\hline  Mexico to Others & 1 \\\hline
Others to Mexico & 0 \\\hline  Others to others & 6 \\ \hline
 \end{tabular}
 \caption{The geographical spread pattern of 2009 A(H1N1). ''Others" refers to other countries except USA and Mexico.  \label{geo} }
\end{center}
\end{table}

\clearpage
\begin{table}
\begin{center}
\begin{tabular}{ | c | c | c | c | c |}
\hline
Position in HA1 protein of H3N2 & 76 & 160 & 172 & 203 \\\hline
Amino acid in consensus strain 1 & Glu & Asn & Lys & Asn \\hline
Amino acid in consensus strain 2 & Lys & Lys & Asn & Lys \\\hline
Shannon Entropy & 0.43 & 0.67 & 0.59 & 0.50 \\ \hline
  \end{tabular}
 \caption{ Consensus strain 1 is the calculated from all strains in the cluster on the right side of Figure 5(c). Consensus strain 2 is the calculated from all strains in the cluster on the left side of Figure 5(c).
  \label{consensus} }
\end{center}
\end{table}

\clearpage

 \setlongtables
\begin{longtable}{ | l | l | l | l | l |}
 \caption{The 2009 A(H1N1)
(swine flu) sequences  in Figure 1. Data are from NCBI
Influenza Virus Resources. \label{sequences1}
 }\\

\hline
 Label & Accession number & Country & Collection date  & Virus name
  \\
  & & & (mm/dd/yyyy) & \\ \hline
 1   &   ACP41926    &   USA     &   3/30/2009   &
A/California/05/2009    \\ \hline 2   &   ACP41105    &   USA     &
4/1/2009    &   A/California/04/2009    \\ \hline 3   &   ACR09364 &
Mexico  &   4/2/2009    &   A/Mexico/4108/2009  \\ \hline 4   &
ACR09372    &   Mexico  &   4/2/2009    &   A/Mexico/3955/2009  \\
\hline 5   &   ACQ99613    &   Mexico  &   4/2/2009    &
A/Mexico/4108/2009  \\ \hline 6   &   ACP44189    &   USA     &
4/9/2009    &   A/California/07/2009    \\ \hline 7   &   ACP41953
&   USA     &   4/9/2009    &   A/California/07/2009    \\ \hline 8
&   ACT36657    &   USA     &   4/14/2009   &   A/Texas/04/2009
\\ \hline 9   &   ACR49285    &   USA     &   4/14/2009   &
A/Texas/04/2009     \\ \hline 10  &   ACU29959    &   USA     &
4/15/2009   &   A/San Antonio/PR922/2009    \\ \hline 11  &
ACQ83310    &   USA     &   4/15/2009   &   A/Texas/15/2009     \\
\hline 12  &   ACP41934    &   USA     &   4/15/2009   &
A/Texas/05/2009     \\ \hline 13  &   ACQ99610    &   Mexico  &
4/19/2009   &   A/Mexico/4603/2009  \\ \hline 14  &   ACR09375    &
Mexico  &   4/20/2009   &   A/Mexico/4635/2009  \\ \hline 15  &
ACR67180    &   USA     &   4/21/2009   &   A/California/13/2009
\\ \hline 16  &   ACT36662    &   USA     &   4/22/2009   &
A/California/12/2009    \\ \hline 17  &   ACQ76383    &   USA     &
4/22/2009   &   A/Indiana/09/2009   \\ \hline 18  &   ACR67176    &
USA     &   4/23/2009   &   A/Texas/10/2009     \\ \hline 19  &
ACU29969    &   USA     &   4/23/2009   &   A/San Antonio/PR923/2009
\\ \hline 20  &   ACR18983    &   USA     &   4/24/2009   &
A/Kansas/03/2009    \\ \hline 21  &   ACR18990    &   USA     &
4/24/2009   &   A/New York/31/2009  \\ \hline 22  &   ACR18980    &
USA     &   4/24/2009   &   A/Texas/08/2009     \\ \hline 23  &
ACQ63286    &   USA     &   4/24/2009   &   A/Ohio/07/2009  \\
\hline 24  &   ACQ73385    &   Canada  &   4/24/2009   &
A/Canada-NS/RV1535/2009     \\ \hline 25  &   ACQ76340    &   USA
&   4/24/2009   &   A/Kansas/03/2009    \\ \hline 26  &   ACQ76386
&   USA     &   4/24/2009   &   A/Ohio/07/2009  \\ \hline 27  &
ACR49292    &   USA     &   4/24/2009   &   A/California/11/2009
\\ \hline 28  &   ACP44147    &   USA     &   4/25/2009   &   A/New
York/19/2009  \\ \hline 29  &   ACT36665    &   USA     &
4/25/2009   &   A/New York/45/2009  \\ \hline 30  &   ACQ63209    &
USA     &   4/25/2009   &   A/New York/12/2009  \\ \hline 31  &
ACR18986    &   USA     &   4/25/2009   &   A/New York/11/2009  \\
\hline 32  &   ACR18987    &   USA     &   4/25/2009   &   A/New
York/09/2009  \\ \hline 33  &   ACR18994    &   USA     &
4/25/2009   &   A/New York/12/2009  \\ \hline 34  &   ACR67173    &
USA     &   4/25/2009   &   A/New York/35/2009  \\ \hline 35  &
ACR67169    &   USA     &   4/26/2009   &   A/South Carolina/10/2009
\\ \hline 36  &   ACQ76333    &   USA     &   4/26/2009   &
A/South Carolina/09/2009    \\ \hline 37  &   ACQ76362    &   USA
&   4/26/2009   &   A/Arizona/02/2009   \\ \hline 38  &   ACR49284
&   USA     &   4/26/2009   &   A/South Carolina/09/2009    \\
\hline 39  &   ACR08534    &   USA     &   4/27/2009   &
A/Texas/23/2009     \\ \hline 40  &   ACR08429    &   USA     &
4/27/2009   &   A/New York/3012/2009    \\ \hline 41  &   ACR08526
&   USA     &   4/27/2009   &   A/Florida/04/2009   \\ \hline 42  &
ACR49289    &   USA     &   4/27/2009   &   A/Arizona/04/2009   \\
\hline 43  &   ACR38825    &   USA     &   4/27/2009   &
A/Georgia/01/2009   \\ \hline 44  &   ACR18991    &   USA     &
4/27/2009   &   A/Washington/12/2009    \\ \hline 45  &   ACT68105
&   Mexico  &   4/27/2009   &   A/Mexico/InDRE13547/2009    \\
\hline 46  &   ACQ84467    &   USA     &   4/27/2009   &   A/New
York/1682/2009    \\ \hline 47  &   ACR67194    &   USA     &
4/28/2009   &   A/Delaware/08/2009  \\ \hline 48  &   ACR15621    &
United Kingdom  &   4/28/2009   &   A/England/195/2009  \\ \hline 49
&   ACR49294    &   USA     &   4/28/2009   &   A/Delaware/05/2009
\\ \hline 50  &   ACS94554    &   USA     &   4/28/2009   &
A/Delaware/06/2009  \\ \hline 51  &   ACS94543    &   USA     &
4/28/2009   &   A/Delaware/07/2009  \\ \hline 52  &   ACR56450    &
USA     &   4/28/2009   &   A/New York/3199/2009    \\ \hline 53  &
ACT33094    &   Germany     &   4/29/2009   &   A/Bayern/62/2009
\\ \hline 54  &   ACT68103    &   Mexico  &   4/29/2009   &
A/Mexico/InDRE13494/2009    \\ \hline 55  &   ACR43957    &   France
&   4/29/2009   &   A/Paris/2573/2009   \\ \hline 56  &   ACR67179
&   USA     &   4/30/2009   &   A/Kentucky/05/2009  \\ \hline 57  &
ACR43939    &   France  &   4/30/2009   &   A/Paris/2590/2009   \\
\hline 58  &   ACR39453    &   USA     &   4/30/2009   &   A/New
York/3232/2009    \\ \hline 59  &   ACR18920    &   Hong Kong   &
4/30/2009   &   A/Hong Kong/01/2009     \\ \hline 60  &   ACT36664
&   USA     &   4/30/2009   &   A/California/19/2009    \\ \hline 61
&   ACT36651    &   USA     &   4/30/2009   &   A/Tennessee/07/2009
\\ \hline 62  &   ACR81633    &   France  &   5/1/2009    &
A/Paris/2591/2009   \\ \hline 63  &   ACS27239    &   USA     &
5/1/2009    &   A/New York/3307/2009    \\ \hline 64  &   ACU13100
&   USA     &   5/1/2009    &   A/California/27/2009    \\ \hline 65
&   ACT36660    &   USA     &   5/1/2009    &   A/California/20/2009
\\ \hline 66  &   ACR67178    &   USA     &   5/4/2009    &
A/Vermont/03/2009   \\ \hline 67  &   ACU13096    &   USA     &
5/5/2009    &   A/Texas/30/2009     \\ \hline 68  &   ACS72650    &
USA     &   5/5/2009    &   A/Michigan/05/2009  \\ \hline 69  &
ACS78026    &   USA     &   5/6/2009    &   A/New York/3351/2009
\\ \hline 70  &   ACS91400    &   France  &   5/6/2009    &
A/Strasbourg/2611/2009  \\ \hline 71  &   ACT68109    &   Canada  &
5/7/2009    &   A/Canada-QC/RV1759/2009     \\ \hline 72  &
ACT68111    &   Canada  &   5/8/2009    &   A/Canada-SK/RV1793/2009
\\ \hline 73  &   ACR56352    &   Sweden  &   5/8/2009    &
A/Stockholm/29/2009     \\ \hline 74  &   ACR32996    &   Finland
&   5/10/2009   &   A/Finland/553/2009  \\ \hline 75  &   ACR38877
&   China   &   5/10/2009   &   A/Shandong/1/2009   \\ \hline 76  &
ACS77996    &   USA     &   5/14/2009   &   A/New York/3463/2009
\\ \hline 77  &   ACT86049    &   USA     &   5/16/2009   &   A/New
York/3502/2009    \\ \hline 78  &   ACR46991    &   Japan   &
5/16/2009   &   A/Osaka/1/2009  \\ \hline 79  &   ACS91402    &
France  &   5/17/2009   &   A/Paris/2650/2009   \\ \hline 80  &
ACS36645    &   China   &   5/18/2009   &   A/GuangzhouSB/01/2009
\\ \hline 81  &   ACS72646    &   USA     &   5/19/2009   &
A/Arizona/09/2009   \\ \hline 82  &   ACT68114    &   Canada  &
5/20/2009   &   A/Canada-MB/RV1964/2009     \\ \hline 83  &
ACT82512    &   Brazil  &   5/20/2009   &   A/Sao Paulo/2056/2009
\\ \hline 84  &   ACS34967    &   Japan   &   5/21/2009   &
A/Sakai/2/2009  \\ \hline 85  &   ACT68117    &   Canada  &
5/21/2009   &   A/Canada-MB/RV1982/2009     \\ \hline 86  &
ACR67262    &   Italy   &   5/21/2009   &   A/Milan/UHSR1/2009  \\
\hline 87  &   ACU29999    &   Taiwan  &   5/22/2009   &
A/Taiwan/T1339/2009     \\ \hline 88  &   ACS34968    &   Japan   &
5/22/2009   &   A/Shiga/1/2009  \\ \hline 89  &   ACS27770    &
Russia  &   5/22/2009   &   A/Kaluga/01/2009    \\ \hline 90  &
ACS91399    &   France  &   5/22/2009   &   A/Paris/2670/2009   \\
\hline 91  &   ACR54964    &   China   &   5/23/2009   &
A/Zhejiang/1/2009   \\ \hline 92  &   ACR67254    &   China   &
5/23/2009   &   A/Beijing/4/2009    \\ \hline 93  &   ACT68101    &
Canada  &   5/24/2009   &   A/Canada-MB/RV2023/2009     \\ \hline 94
&   ACR78588    &   Russia  &   5/26/2009   &   A/Moscow/IIV02/2009
\\ \hline 95  &   ACS50088    &   Finland     &   5/26/2009   &
A/Finland/554/2009  \\ \hline 96  &   ACT33123    &   Brazil  &
5/28/2009   &   A/Mato Grosso/2329/2009     \\ \hline 97  &
ACU30009    &   Taiwan  &   5/28/2009   &   A/Taiwan/T1773/2009
\\ \hline 98  &   ACT68118    &   Canada  &   5/28/2009   &
A/Canada-MB/RV2013/2009     \\ \hline 99  &   ACT79625    &   China
&   5/28/2009   &   A/Shanghai/60T/2009     \\ \hline 100 &
ACT82514    &   Brazil  &   5/29/2009   &   A/Sao Paulo/2261/2009
\\ \hline 101 &   ACR83538    &   China   &   5/29/2009   &
A/Guangdong/03/2009     \\ \hline 102 &   ACT21932    &   Sweden  &
5/29/2009   &   A/Stockholm/31/2009     \\ \hline 103 &   ACS68822
&   China   &   5/31/2009   &   A/Zhejiang/2/2009   \\ \hline 104 &
ACT21971    &   Sweden  &   6/1/2009    &   A/Stockholm/35/2009
\\ \hline 105 &   ACS94551    &   USA     &   6/1/2009    &
A/Rhode Island/04/2009  \\ \hline 106 &   ACS54262    &   Japan   &
6/2/2009    &   A/Tokushima/1/2009  \\ \hline 107 &   ACS91398    &
France  &   6/2/2009    &   A/Paris/2722/2009   \\ \hline 108 &
BAH95823    &   Japan   &   6/3/2009    &   A/Saitama/55/2009   \\
\hline 109 &   ACS45035    &   USA     &   6/5/2009    &
A/California/04/2009    \\ \hline 110 &   ACT79624    &   China   &
6/6/2009    &   A/Shanghai/143T/2009    \\ \hline 111 &   ACT83737
&   Italy   &   6/8/2009    &   A/Ancona/01/2009    \\ \hline 112 &
ACT10316    &   Hong Kong   &   6/11/2009   &   A/Hong
Kong/2369/2009   \\ \hline 113 &   ACT83738    &   Italy   &
6/11/2009   &   A/Ancona/02/2009    \\ \hline 114 &   ACS92600    &
Japan   &   6/16/2009   &   A/Utsunomiya/1/2009     \\ \hline 115 &
ACU17389    &   USA     &   6/17/2009   &   A/New York/4197/2009
\\ \hline 116 &   ACT79183    &   USA     &   6/18/2009   &
A/Bethesda/SP508/2009   \\ \hline 117 &   ACS45017    &   China   &
6/18/2009   &   A/Nanjing/1/2009    \\ \hline 118 &   ACU30112    &
USA     &   6/18/2009   &   A/Silver Spring/SP509/2009  \\ \hline
119 &   ACU30121    &   USA     &   6/18/2009   &   A/Silver
Spring/SP510/2009  \\ \hline 120 &   ACT22055    &   Chile   &
6/19/2009   &   A/Puerto Montt/Bio87/2009   \\ \hline 121 &
ACU30073    &   Colombia    &   6/25/2009   &   A/Bogota/0466N/2009
\\ \hline 122 &   ACU13129    &   China   &   6/29/2009   &
A/Changsha/78/2009  \\ \hline 123 &   ACT79133    &   Japan   &
6/29/2009   &   A/Japan/1070/2009   \\ \hline 124 &   ACT66162    &
Singapore   &   6/30/2009   &   A/Singapore/TLL01/2009  \\ \hline
125 &   ACT83739    &   Italy   &   7/1/2009    &   A/Ancona/04/2009
\\ \hline 126 &   ACT82516    &   Brazil  &   7/3/2009    &   A/Sao
Paulo/43812/2009  \\ \hline 127 &   ACT83741    &   Italy   &
7/12/2009   &   A/Ancona/05/2009    \\ \hline

\end{longtable}
\clearpage

 \setlongtables
\begin{longtable}{ | l | l | l | l | l |}
 \caption{The H3N2 sequences in Figure 4(c). Data are from NCBI
Influenza Virus Resources. \label{sequences4c}
 }\\

\hline
 Label & Accession number & Country & Collection date & Virus name
  \\  & & & (mm/dd/yyyy) & \\ \hline
1	&	ACF36384 	&	Hong Kong 	&	10/1/2002	&	A/Hong Kong/CUHK50200/2002	\\ \hline
2	&	ACC66393 	&	Australia 	&	10/2/2002	&	A/VICTORIA/432/2002	\\ \hline
3	&	ABI92632 	&	Australia 	&	10/15/2002	&	A/Western Australia/36/2002	\\ \hline
4	&	ACC67637 	&	France 	&	10/15/2002	&	A/LYON/19989/2002	\\ \hline
5	&	ACC77932 	&	USA 	&	10/30/2002	&	A/Hawaii/HI-02-3736/2002	\\ \hline
6	&	ACC66357 	&	South Korea 	&	10/31/2002	&	A/Kwangju/219/2002	\\ \hline
7	&	ACC77933 	&	USA 	&	11/1/2002	&	A/New York/28/2002	\\ \hline
8	&	ACC77934 	&	Bulgaria 	&	11/2/2002	&	A/Sofia/343/2002	\\ \hline
9	&	ACC67720 	&	Bulgaria 	&	11/2/2002	&	A/SOFIA/344/2002	\\ \hline
10	&	ACC66354 	&	South Korea 	&	11/14/2002	&	A/Incheon/260/2002	\\ \hline
11	&	ACC66358 	&	South Korea 	&	11/18/2002	&	A/KYONGBUK/304/2002	\\ \hline
12	&	ACC66369 	&	South Korea 	&	11/20/2002	&	A/Pusan/504/2002	\\ \hline
13	&	ACC66345 	&	South Korea 	&	11/20/2002	&	A/Cheonnam/323/2002	\\ \hline
14	&	ACC77936 	&	China 	&	11/21/2002	&	A/Beijing/178-NEW-AGN/2002	\\ \hline
15	&	ACC77937 	&	Philippines 	&	11/21/2002	&	A/Philippines/PH-1159050/2002	\\ \hline
16	&	ACC66344 	&	South Korea 	&	11/29/2002	&	A/Cheju/274/2002	\\ \hline
17	&	ACC77939 	&	China 	&	12/1/2002	&	A/Anhui/550/2002	\\ \hline
18	&	ACC67682 	&	France 	&	12/1/2002	&	A/PARIS/207/2002	\\ \hline
19	&	ACC77940 	&	USA 	&	12/2/2002	&	A/Alaska/AK-RSP-02-753/2002	\\ \hline
20	&	ACN32523 	&	Italy 	&	12/2/2002	&	A/Genoa/1/2002	\\ \hline
21	&	ABO37625 	&	South Korea 	&	12/2/2002	&	A/Korea/770/2002	\\ \hline
22	&	ACC77942 	&	USA 	&	12/3/2002	&	A/California/CA-T02-3025/2002	\\ \hline
23	&	ACC66374 	&	Singapore 	&	12/7/2002	&	A/SINGAPORE/44/2002	\\ \hline
24	&	ACC77944 	&	Taiwan 	&	12/9/2002	&	A/Taiwan/TW-1521/2002	\\ \hline
25	&	ACC77945 	&	Turkey 	&	12/13/2002	&	A/Turkey/TU-RSP-02-1035/2002	\\ \hline
26	&	ACC77946 	&	Taiwan 	&	12/15/2002	&	A/Taiwan/TW-1522/2002	\\ \hline
27	&	ACF36415 	&	Hong Kong 	&	12/18/2002	&	A/Hong Kong/CUHK53123/2002	\\ \hline
28	&	ACC77948 	&	China 	&	12/20/2002	&	A/Beijing/301/2002	\\ \hline
29	&	ACC77949 	&	USA 	&	12/22/2002	&	A/Hawaii/HI-02-4051/2002	\\ \hline
30	&	ACC67657 	&	Latvia 	&	12/23/2002	&	A/LATVIA/13794/2002	\\ \hline
31	&	ACC77950 	&	USA 	&	12/24/2002	&	A/Hawaii/HI-02-4055/2002	\\ \hline
32	&	ABO10182 	&	Japan 	&	12/25/2002	&	A/Kumamoto/102/2002E	\\ \hline
33	&	ACC77951 	&	USA 	&	12/28/2002	&	A/Hawaii/HI-02-4092/2002	\\ \hline
34	&	ACC77952 	&	USA 	&	12/29/2002	&	A/North Carolina/NC-C02-4981/2002	\\ \hline
35	&	ACC66516 	&	Singapore 	&	1/7/2003	&	A/SINGAPORE/7/2003	\\ \hline
36	&	ACC66445 	&	United Kingdom 	&	1/8/2003	&	A/ENGLAND/1/2003	\\ \hline
37	&	ACC67602 	&	USA 	&	1/8/2003	&	A/Washington/WA-V130811/2003	\\ \hline
38	&	ACC67661 	&	Germany 	&	1/14/2003	&	A/BAYERN/1/2003	\\ \hline
39	&	ACC67377 	&	Egypt 	&	1/14/2003	&	A/Egypt/EG-2002923226/2003	\\ \hline
40	&	ACC67440 	&	USA 	&	1/15/2003	&	A/Michigan/MI-VC76/2003	\\ \hline
41	&	ACC66517 	&	Singapore 	&	1/15/2003	&	A/SINGAPORE/16/2003	\\ \hline
42	&	ABO20960 	&	Australia 	&	1/15/2003	&	A/Sydney/015/03	\\ \hline
43	&	ACC66485 	&	USA 	&	1/16/2003	&	A/MEMPHIS/1/2003	\\ \hline
44	&	ACC66486 	&	USA 	&	1/16/2003	&	A/Memphis/2/2003	\\ \hline
45	&	ACF36424 	&	Hong Kong 	&	1/18/2003	&	A/Hong Kong/CUHK5627/2003	\\ \hline
46	&	ACF36426 	&	Hong Kong 	&	1/20/2003	&	A/Hong Kong/CUHK5723/2003	\\ \hline
47	&	ABB03112 	&	USA 	&	1/20/2003	&	A/New York/485/2003	\\ \hline
48	&	ACC67688 	&	Ireland 	&	1/24/2003	&	A/IRELAND/1215/2003	\\ \hline
49	&	ACC66560 	&	USA 	&	1/28/2003	&	A/WYOMING/2/2003	\\ \hline
50	&	ACC66521 	&	Bulgaria 	&	2/1/2003	&	A/SOFIA/141/2003	\\ \hline

\end{longtable}
\clearpage
\clearpage

 \setlongtables
\begin{longtable}{ | l | l | l | l | l |}
 \caption{The H3N2 sequences in Figure 4(d). Data are from NCBI
Influenza Virus Resources. \label{sequences4d}
 }\\

\hline
 Label & Accession number & Country & Collection date & Virus name
  \\  & & & (mm/dd/yyyy) & \\ \hline

  1	&	ABI92544 	&	Australia 	&	10/2/2001	&	A/Western Australia/17/2001	\\ \hline	
2	&	ACC66320 	&	Australia 	&	10/29/2001	&	A/South Australia/102/2001	\\ \hline	
3	&	ACF36383 	&	Hong Kong 	&	11/8/2001	&	A/Hong Kong/CUHK50080/2001	\\ \hline	
4	&	AAX56570 	&	USA 	&	11/22/2001	&	A/New York/71/2001	\\ \hline	
5	&	AAX56490 	&	USA 	&	11/27/2001	&	A/New York/124/2001	\\ \hline	
6	&	AAX11625 	&	USA 	&	12/15/2001	&	A/New York/83/2001	\\ \hline	
7	&	AAX35861 	&	USA 	&	12/23/2001	&	A/New York/131/2001	\\ \hline	
8	&	AAY28335 	&	USA 	&	12/25/2001	&	A/New York/94/2001	\\ \hline	
9	&	ACF36396 	&	Hong Kong 	&	12/26/2001	&	A/Hong Kong/CUHK51424/2001	\\ \hline	
10	&	ACF36397 	&	Hong Kong 	&	12/26/2001	&	A/Hong Kong/CUHK51431/2001	\\ \hline	
11	&	ACF36398 	&	Hong Kong 	&	12/28/2001	&	A/Hong Kong/CUHK51490/2001	\\ \hline	
12	&	AAX12761 	&	USA 	&	12/29/2001	&	A/New York/84/2001	\\ \hline	
13	&	AAX56440 	&	USA 	&	12/31/2001	&	A/New York/85/2001	\\ \hline	
14	&	ACC77886 	&	China 	&	1/1/2002	&	A/Tianjin/5/2002	\\ \hline	
15	&	ACC77887 	&	USA 	&	1/1/2002	&	A/Louisiana/2/2002	\\ \hline	
16	&	ACC77888 	&	USA 	&	1/4/2002	&	A/Oklahoma/2/2002	\\ \hline	
17	&	ABA42335 	&	USA 	&	1/4/2002	&	A/New York/403/2002	\\ \hline	
18	&	ACC77889 	&	USA 	&	1/5/2002	&	A/Utah/6/2002	\\ \hline	
19	&	ACC77891 	&	China 	&	1/7/2002	&	A/Beijing/20/2002	\\ \hline	
20	&	ABA42989 	&	USA 	&	1/7/2002	&	A/New York/405/2002	\\ \hline	
21	&	ACC66370 	&	Singapore 	&	1/7/2002	&	A/SINGAPORE/2/2002	\\ \hline	
22	&	ACC66386 	&	Australia 	&	1/9/2002	&	A/VICTORIA/103/2002	\\ \hline	
23	&	ACC66387 	&	Australia 	&	1/9/2002	&	A/VICTORIA/105/2002	\\ \hline	
24	&	ACF36405 	&	Hong Kong 	&	1/10/2002	&	A/Hong Kong/CUHK5250/2002	\\ \hline	
25	&	ACF36406 	&	Hong Kong 	&	1/10/2002	&	A/Hong Kong/CUHK5251/2002	\\ \hline	
26	&	ACC77895 	&	China 	&	1/10/2002	&	A/Wuhan/12/2002	\\ \hline	
27	&	ABA42368 	&	USA 	&	1/11/2002	&	A/New York/408/2002	\\ \hline	
28	&	ACC66544 	&	Australia 	&	1/12/2002	&	A/VICTORIA/102/2003	\\ \hline	
29	&	ACF36411 	&	Hong Kong 	&	1/12/2002	&	A/Hong Kong/CUHK5296/2002	\\ \hline	
30	&	AAX56460 	&	USA 	&	1/12/2002	&	A/New York/89/2002	\\ \hline	
31	&	ABR15962 	&	New Zealand 	&	1/14/2002	&	A/Auckland/614/2002	\\ \hline	
32	&	ABA42379 	&	USA 	&	1/14/2002	&	A/New York/409/2002	\\ \hline	
33	&	AAY47023 	&	USA 	&	1/14/2002	&	A/New York/125/2002	\\ \hline	
34	&	ABA42487 	&	USA 	&	1/16/2002	&	A/New York/418/2002	\\ \hline	
35	&	AAY28648 	&	USA 	&	1/16/2002	&	A/New York/107/2002	\\ \hline	
36	&	ACC77896 	&	China 	&	1/17/2002	&	A/Wuhan/16/2002	\\ \hline	
37	&	ABR14636 	&	Norway 	&	1/17/2002	&	A/Oslo/398/2002	\\ \hline	
38	&	AAX12771 	&	USA 	&	1/17/2002	&	A/New York/91/2002	\\ \hline	
39	&	AAX47525 	&	USA 	&	1/18/2002	&	A/New York/96/2002	\\ \hline	
40	&	ABA42498 	&	USA 	&	1/18/2002	&	A/New York/419/2002	\\ \hline	
41	&	ACC77897 	&	Hong Kong 	&	1/19/2002	&	A/Hong Kong/1550/2002	\\ \hline	
42	&	AAX56580 	&	USA 	&	1/20/2002	&	A/New York/106/2002	\\ \hline	
43	&	AAX12791 	&	USA 	&	1/22/2002	&	A/New York/100/2002	\\ \hline	
44	&	ABA42401 	&	USA 	&	1/22/2002	&	A/New York/411/2002	\\ \hline	
45	&	ACC77898 	&	USA 	&	1/23/2002	&	A/New York/16/2002	\\ \hline	
46	&	ACC66371 	&	Singapore 	&	1/24/2002	&	A/SINGAPORE/7/2002	\\ \hline	
47	&	AAX56510 	&	USA 	&	1/27/2002	&	A/New York/132/2002	\\ \hline	
48	&	ABB04983 	&	USA 	&	1/28/2002	&	A/New York/420/2002	\\ \hline	
49	&	AAX11635 	&	USA 	&	1/29/2002	&	A/New York/110/2002	\\ \hline	
50	&	AAY44641 	&	USA 	&	1/30/2002	&	A/New York/118/2002	\\ \hline	
51	&	ABG48082 	&	New Zealand 	&	1/31/2002	&	A/Waikato/2/2002	\\ \hline	
52	&	AAX56590 	&	USA 	&	1/31/2002	&	A/New York/108/2002	\\ \hline	
53	&	AAX12801 	&	USA 	&	1/31/2002	&	A/New York/113/2002	\\ \hline	
54	&	ABA42412 	&	USA 	&	2/1/2002	&	A/New York/412/2002	\\ \hline	
55	&	ABI21508 	&	USA 	&	2/1/2002	&	A/New York/C4/2002	\\ \hline	
56	&	ACC77901 	&	USA 	&	2/1/2002	&	A/Massachusetts/2/2002	\\ \hline	
57	&	ACC77902 	&	USA 	&	2/4/2002	&	A/New York/18/2002	\\ \hline	
58	&	AAY28014 	&	USA 	&	2/4/2002	&	A/New York/101/2002	\\ \hline	
59	&	AAX56470 	&	USA 	&	2/6/2002	&	A/New York/90/2002	\\ \hline	
60	&	ABG48093 	&	New Zealand 	&	2/10/2002	&	A/Wellington/6/2002	\\ \hline	
61	&	AAY44620 	&	USA 	&	2/12/2002	&	A/New York/75/2002	\\ \hline	
62	&	AAX47515 	&	USA 	&	2/12/2002	&	A/New York/129/2002	\\ \hline	
63	&	AAX35871 	&	USA 	&	2/14/2002	&	A/New York/135/2002	\\ \hline	
64	&	ACC66327 	&	Australia 	&	2/18/2002	&	A/BRISBANE/3/2002	\\ \hline	
65	&	AAX57784 	&	USA 	&	2/19/2002	&	A/New York/120/2002	\\ \hline	
66	&	ACC77904 	&	USA 	&	2/21/2002	&	A/North Carolina/6/2002	\\ \hline	
67	&	ACC66353 	&	Japan 	&	2/23/2002	&	A/FUKUOKA/15/2002	\\ \hline	
68	&	ACC77905 	&	USA 	&	2/24/2002	&	A/Illinois/1/2002	\\ \hline	
69	&	BAG28225 	&	Japan 	&	2/25/2002	&	A/Morioka/34/2002	\\ \hline	
70	&	AAY28325 	&	USA 	&	2/26/2002	&	A/New York/88/2002	\\ \hline	
71	&	AAX12781 	&	USA 	&	2/26/2002	&	A/New York/95/2002	\\ \hline	
72	&	ABA42476 	&	USA 	&	2/28/2002	&	A/New York/416/2002	\\ \hline	
73	&	AAX35841 	&	USA 	&	3/1/2002	&	A/New York/86/2002	\\ \hline	
74	&	ACF36221 	&	Hong Kong 	&	3/1/2002	&	A/Hong Kong/CUHK13216/2002	\\ \hline	
75	&	BAG28226 	&	Japan 	&	3/1/2002	&	A/Morioka/52/2002	\\ \hline	
76	&	AAZ38539 	&	USA 	&	3/1/2002	&	A/New York/126/2002	\\ \hline	
77	&	ACF36222 	&	Hong Kong 	&	3/3/2002	&	A/Hong Kong/CUHK13249/2002	\\ \hline	
78	&	ACF36223 	&	Hong Kong 	&	3/4/2002	&	A/Hong Kong/CUHK13278/2002	\\ \hline	
79	&	ABS50064 	&	New Zealand 	&	3/4/2002	&	A/Auckland/615/2002	\\ \hline	
80	&	ACC77909 	&	Thailand 	&	3/4/2002	&	A/Thailand/128400/2002	\\ \hline	
81	&	ACC77910 	&	USA 	&	3/5/2002	&	A/Hawaii/4/2002	\\ \hline	
82	&	ACC77911 	&	Hong Kong 	&	3/6/2002	&	A/Hong Kong/121715/2002	\\ \hline	
83	&	AAY98077 	&	USA 	&	3/7/2002	&	A/New York/276/2002	\\ \hline	
84	&	ACC66376 	&	Thailand 	&	3/8/2002	&	A/SONGKHLA/107/2002	\\ \hline	
85	&	AAX57734 	&	USA 	&	3/11/2002	&	A/New York/130/2002	\\ \hline	
86	&	AAY28405 	&	USA 	&	3/21/2002	&	A/New York/122/2002	\\ \hline	
87	&	ACC77912 	&	Hong Kong 	&	3/25/2002	&	A/Hong Kong/568/2002	\\ \hline	
88	&	ACC66323 	&	Thailand 	&	3/25/2002	&	A/BANGKOK/109/2002	\\ \hline	
89	&	ACC77913 	&	USA 	&	4/3/2002	&	A/Washington/3/2002	\\ \hline	
90	&	ACC77914 	&	Brazil 	&	4/4/2002	&	A/Ceara/177/2002	\\ \hline	
91	&	ACF36278 	&	Hong Kong 	&	4/4/2002	&	A/Hong Kong/CUHK21713/2002	\\ \hline	
92	&	ACC66361 	&	New Caledonia 	&	4/4/2002	&	A/New Caledonia/6/2002	\\ \hline	
93	&	ACF36281 	&	Hong Kong 	&	4/8/2002	&	A/Hong Kong/CUHK21742/2002	\\ \hline	
94	&	ACF36284 	&	Hong Kong 	&	4/11/2002	&	A/Hong Kong/CUHK21957/2002	\\ \hline	
95	&	ABG37417 	&	New Zealand 	&	4/12/2002	&	A/Waikato/5/2002	\\ \hline	
96	&	ACC77915 	&	USA 	&	4/17/2002	&	A/Nebraska/12/2002	\\ \hline	
97	&	ABC42871 	&	New Zealand 	&	4/26/2002	&	A/Canterbury/01/2002	\\ \hline	
98	&	ABC67850 	&	New Zealand 	&	5/2/2002	&	A/Canterbury/14/2002	\\ \hline	
99	&	ACF36299 	&	Hong Kong 	&	5/10/2002	&	A/Hong Kong/CUHK23113/2002	\\ \hline	
100	&	ACF36300 	&	Hong Kong 	&	5/10/2002	&	A/Hong Kong/CUHK23162/2002	\\ \hline	
101	&	ACF36301 	&	Hong Kong 	&	5/12/2002	&	A/Hong Kong/CUHK23180/2002	\\ \hline	
102	&	ACC77916 	&	Argentina 	&	5/14/2002	&	A/Argentina/1/2002	\\ \hline	
103	&	ACC77917 	&	Peru 	&	5/14/2002	&	A/Peru/3029/2002	\\ \hline	
104	&	ACC77918 	&	China 	&	5/14/2002	&	A/Guangzhou/394/2002	\\ \hline	
105	&	ACC66394 	&	Australia 	&	5/20/2002	&	A/VICTORIA/506/2002	\\ \hline	
106	&	ACC77919 	&	Brazil 	&	5/23/2002	&	A/Brazil/1727/2002	\\ \hline	
107	&	ACF36302 	&	Hong Kong 	&	6/1/2002	&	A/Hong Kong/CUHK24044/2002	\\ \hline	
108	&	ACF36303 	&	Hong Kong 	&	6/2/2002	&	A/Hong Kong/CUHK24054/2002	\\ \hline	
109	&	ACF36304 	&	Hong Kong 	&	6/3/2002	&	A/Hong Kong/CUHK24114/2002	\\ \hline	
110	&	ACF36305 	&	Hong Kong 	&	6/4/2002	&	A/Hong Kong/CUHK24167/2002	\\ \hline	
111	&	ACC77920 	&	Hong Kong 	&	6/6/2002	&	A/Hong Kong/1143/2002	\\ \hline	
112	&	ABI92335 	&	New Zealand 	&	6/7/2002	&	A/Dunedin/1/2002	\\ \hline	
113	&	ACC66328 	&	Australia 	&	6/7/2002	&	A/BRISBANE/5/2002	\\ \hline	
114	&	ABC50255 	&	New Zealand 	&	6/8/2002	&	A/Canterbury/47/2002	\\ \hline	
115	&	ABR28856 	&	New Zealand 	&	6/8/2002	&	A/Auckland/608/2002	\\ \hline	
116	&	ABC50299 	&	New Zealand 	&	6/10/2002	&	A/Canterbury/53/2002	\\ \hline	
117	&	ABC50288 	&	New Zealand 	&	6/11/2002	&	A/Canterbury/50/2002	\\ \hline	
118	&	ACC66347 	&	New Zealand 	&	6/13/2002	&	A/CHRISTCHURCH/37/2002	\\ \hline	
119	&	ABC68049 	&	New Zealand 	&	6/13/2002	&	A/South Canterbury/37/2002	\\ \hline	
120	&	ABG37428 	&	New Zealand 	&	6/17/2002	&	A/Wellington/38/2002	\\ \hline	
121	&	ABC67989 	&	New Zealand 	&	6/20/2002	&	A/Canterbury/33/2002	\\ \hline	
122	&	ACC66329 	&	Australia 	&	6/20/2002	&	A/BRISBANE/6/2002	\\ \hline	
123	&	ABG88806 	&	Australia 	&	6/25/2002	&	A/Western Australia/23/2002	\\ \hline	
124	&	ABC67653 	&	New Zealand 	&	6/27/2002	&	A/Canterbury/57/2002	\\ \hline	
125	&	ACC77922 	&	Brazil 	&	6/27/2002	&	A/Brazil/1732/2002	\\ \hline	
126	&	ABC85952 	&	New Zealand 	&	6/30/2002	&	A/Canterbury/60/2002	\\ \hline	
127	&	ABG37450 	&	New Zealand 	&	7/1/2002	&	A/Waikato/21/2002	\\ \hline	
128	&	ACC66324 	&	Thailand 	&	7/1/2002	&	A/BANGKOK/190/2002	\\ \hline	
129	&	ACF36339 	&	Hong Kong 	&	7/1/2002	&	A/Hong Kong/CUHK33047/2002	\\ \hline	
130	&	ABC50321 	&	New Zealand 	&	7/2/2002	&	A/Canterbury/59/2002	\\ \hline	
131	&	ACF36340 	&	Hong Kong 	&	7/2/2002	&	A/Hong Kong/CUHK33079/2002	\\ \hline	
132	&	ACF36341 	&	Hong Kong 	&	7/2/2002	&	A/Hong Kong/CUHK33106/2002	\\ \hline	
133	&	ACC66373 	&	Singapore 	&	7/2/2002	&	A/SINGAPORE/29/2002	\\ \hline	
134	&	ABI92346 	&	New Zealand 	&	7/3/2002	&	A/Waikato/23/2002	\\ \hline	
135	&	ACC66381 	&	Australia 	&	7/4/2002	&	A/SYDNEY/23/2002	\\ \hline	
136	&	ABC68093 	&	New Zealand 	&	7/6/2002	&	A/Canterbury/72/2002	\\ \hline	
137	&	ABC85919 	&	New Zealand 	&	7/6/2002	&	A/Canterbury/70/2002	\\ \hline	
138	&	ACC66380 	&	Australia 	&	7/7/2002	&	A/SYDNEY/21/2002	\\ \hline	
139	&	ABG26956 	&	New Zealand 	&	7/9/2002	&	A/Dunedin/10/2002	\\ \hline	
140	&	ACC66321 	&	New Zealand 	&	7/9/2002	&	A/AUCKLAND/26/2002	\\ \hline	
141	&	ABC67543 	&	New Zealand 	&	7/9/2002	&	A/Canterbury/79/2002	\\ \hline	
142	&	ABC50376 	&	New Zealand 	&	7/9/2002	&	A/Canterbury/81/2002	\\ \hline	
143	&	ABI92555 	&	Australia 	&	7/10/2002	&	A/Western Australia/25/2002	\\ \hline	
144	&	ABC50354 	&	New Zealand 	&	7/11/2002	&	A/Canterbury/75/2002	\\ \hline	
145	&	ABC50365 	&	New Zealand 	&	7/11/2002	&	A/Canterbury/80/2002	\\ \hline	
146	&	ABG48137 	&	New Zealand 	&	7/15/2002	&	A/Wellington/71/2002	\\ \hline	
147	&	ACC66330 	&	Australia 	&	7/15/2002	&	A/BRISBANE/22/2002	\\ \hline	
148	&	ACC66331 	&	Australia 	&	7/17/2002	&	A/BRISBANE/45/2002	\\ \hline	
149	&	ACC77923 	&	Philippines 	&	7/17/2002	&	A/Philippines/160283/2002	\\ \hline	
150	&	ACC66367 	&	Philippines 	&	7/18/2002	&	A/PHILIPPINES/471/2002	\\ \hline	
151	&	ABK39951 	&	Australia 	&	7/21/2002	&	A/Western Australia/27/2002	\\ \hline	
152	&	ABG88630 	&	New Zealand 	&	7/23/2002	&	A/Waikato/36/2002	\\ \hline	
153	&	ACC77926 	&	Hong Kong 	&	7/26/2002	&	A/Hong Kong/1510/2002	\\ \hline	
154	&	ABI92577 	&	Australia 	&	8/6/2002	&	A/Western Australia/28/2002	\\ \hline	
155	&	ACC66335 	&	Australia 	&	8/8/2002	&	A/BRISBANE/144/2002	\\ \hline	
156	&	ACC77928 	&	Brazil 	&	8/9/2002	&	A/Brazil/2458/2002	\\ \hline	
157	&	ABH01014 	&	China 	&	8/11/2002	&	A/Fujian/411/2002	\\ \hline	
158	&	ACC66336 	&	Australia 	&	8/11/2002	&	A/BRISBANE/157/2002	\\ \hline	
159	&	ACC66352 	&	China 	&	8/11/2002	&	A/Fujian/411/2002	\\ \hline	
160	&	ABP51971 	&	China 	&	8/11/2002	&	A/Fujian/411/2002	\\ \hline	
161	&	ABG88641 	&	New Zealand 	&	8/12/2002	&	A/Dunedin/18/2002	\\ \hline	
162	&	ABK80179 	&	Australia 	&	8/13/2002	&	A/Queensland/23/2002	\\ \hline	
163	&	ABL67308 	&	Australia 	&	8/13/2002	&	A/Queensland/27/2002	\\ \hline	
164	&	ACC66338 	&	Australia 	&	8/13/2002	&	A/BRISBANE/192/2002	\\ \hline	
165	&	ACC77929 	&	India 	&	8/16/2002	&	A/India/25502/2002	\\ \hline	
166	&	ABI92599 	&	Australia 	&	8/19/2002	&	A/Western Australia/30/2002	\\ \hline	
167	&	ABG48170 	&	New Zealand 	&	8/20/2002	&	A/Wellington/79/2002	\\ \hline	
168	&	ACC66364 	&	Australia 	&	8/21/2002	&	A/PERTH/49/2002	\\ \hline	
169	&	ACC66391 	&	Australia 	&	8/22/2002	&	A/VICTORIA/235/2002	\\ \hline	
170	&	ABG37472 	&	New Zealand 	&	8/23/2002	&	A/Waikato/58/2002	\\ \hline	
171	&	ABN51154 	&	Australia 	&	8/28/2002	&	A/Western Australia/34/2002	\\ \hline	
172	&	ACC66322 	&	New Zealand 	&	9/3/2002	&	A/AUCKLAND/57/2002	\\ \hline	
173	&	ACC66342 	&	Australia 	&	9/4/2002	&	A/BRISBANE/312/2002	\\ \hline	
174	&	ABR15940 	&	New Zealand 	&	9/9/2002	&	A/Auckland/611/2002	\\ \hline	
175	&	ACC66360 	&	Malaysia 	&	9/9/2002	&	A/MALAYSIA/145/2002	\\ \hline	
176	&	ACC66379 	&	Australia 	&	9/15/2002	&	A/South Australia/154/2002	\\ \hline	
177	&	ACC66392 	&	Australia 	&	9/16/2002	&	A/VICTORIA/254/2002	\\ \hline	
178	&	ACC66383 	&	Taiwan 	&	9/25/2002	&	A/TAIWAN/8/2002	\\ \hline	
179	&	ABI92621 	&	Australia 	&	9/29/2002	&	A/Western Australia/35/2002	\\ \hline	
180	&	ACC66365 	&	Australia 	&	9/29/2002	&	A/PERTH/89/2002	\\ \hline

  \end{longtable}
\clearpage

\setlongtables
\begin{longtable}{ | l | l | l | l | l |}
 \caption{The sequences of H3N2
viruses after 10/01/2008. Data are from GISAID. The first column
label is the same as used in Figure 5. }\\
 \hline Label
& EPI  ID & Country &   Collection date &  Virus name
\\ & & & (mm/dd/yyyy) & \\ \hline 1

1	&	EPI187670	&	Japan 	&	4/2/2009	&	A/AICHI/158/2009	\\ \hline
2	&	EPI187671	&	Japan 	&	4/9/2009	&	A/AICHI/161/2009	\\ \hline
3	&	EPI169724	&	Japan 	&	11/26/2008	&	A/AKITA/12/2008	\\ \hline
4	&	EPI187672	&	Japan 	&	5/27/2009	&	A/AKITA/34/2009	\\ \hline
5	&	EPI185826	&	USA 	&	2/27/2009	&	A/Arizona/08/2009	\\ \hline
6	&	EPI185829	&	USA 	&	2/25/2009	&	A/Arizona/11/2009	\\ \hline
7	&	EPI185805	&	Russia 	&	3/23/2009	&	A/Astrakhan/7/2009	\\ \hline
8	&	EPI185723	&	China	&	2/10/2009	&	A/Beijng-Xicheng/1108/2009	\\ \hline
9	&	EPI189216	&	Australia 	&	4/28/2009	&	A/Brisbane/53/2009	\\ \hline
10	&	EPI185704	&	Canada 	&	3/15/2009	&	A/British Columbia/RV1222/2009	\\ \hline
11	&	EPI185707	&	Canada 	&	3/15/2009	&	A/British Columbia/RV1223/2009	\\ \hline
12	&	EPI187312	&	Japan 	&	5/1/2009	&	A/CHIBA-C/42/2009	\\ \hline
13	&	EPI187313	&	Japan 	&	1/16/2009	&	A/CHIBA-C/7/2009	\\ \hline
14	&	EPI175111	&	USA 	&	12/20/2008	&	A/California/10/2008	\\ \hline
15	&	EPI185831	&	USA 	&	1/1/2009	&	A/Colorado/01/2009	\\ \hline
16	&	EPI169244	&	USA 	&	10/15/2008	&	A/Colorado/03/2008	\\ \hline
17	&	EPI172436	&	USA 	&	10/29/2008	&	A/Colorado/04/2008	\\ \hline
18	&	EPI185834	&	USA 	&	1/16/2009	&	A/Colorado/05/2009	\\ \hline
19	&	EPI185837	&	USA 	&	1/21/2009	&	A/Colorado/06/2009	\\ \hline
20	&	EPI175119	&	USA 	&	12/17/2008	&	A/Colorado/12/2008	\\ \hline
21	&	EPI175120	&	USA 	&	12/24/2008	&	A/Colorado/15/2008	\\ \hline
22	&	EPI187314	&	Japan 	&	12/15/2008	&	A/EHIME/36/2008	\\ \hline
23	&	EPI187315	&	Japan 	&	1/19/2009	&	A/EHIME/9/2009	\\ \hline
24	&	EPI187317	&	Japan	&	12/26/2008	&	A/FUKUOKA-C/39/2008	\\ \hline
25	&	EPI187319	&	Japan 	&	12/26/2008	&	A/FUKUSHIMA/124/2008	\\ \hline
26	&	EPI185726	&	China	&	11/26/2008	&	A/Fujian-Siming/1242/2008	\\ \hline
27	&	EPI185729	&	China	&	3/3/2009	&	A/Fujian-Tongan/142/2009	\\ \hline
28	&	EPI187320	&	Japan 	&	1/28/2009	&	A/GIFU-C/37/2009	\\ \hline
29	&	EPI187321	&	Japan 	&	1/28/2009	&	A/GIFU-C/38/2009	\\ \hline
30	&	EPI172442	&	Guam	&	10/27/2008	&	A/Guam/7124/2008	\\ \hline
31	&	EPI187322	&	Japan	&	2/3/2009	&	A/HIROSHIMA-C/20/2009	\\ \hline
32	&	EPI187323	&	Japan	&	1/6/2009	&	A/HIROSHIMA-C/7/2009	\\ \hline
33	&	EPI187674	&	Japan 	&	5/26/2009	&	A/HIROSHIMA/148/2009	\\ \hline
34	&	EPI187675	&	Japan 	&	6/10/2009	&	A/HIROSHIMA/154/2009	\\ \hline
35	&	EPI187324	&	Japan 	&	12/6/2008	&	A/HOKKAIDO/9/2008	\\ \hline
36	&	EPI187325	&	Japan	&	11/29/2008	&	A/HYOGO/6/2008	\\ \hline
37	&	EPI187326	&	Japan	&	12/11/2008	&	A/HYOGO/99/2008	\\ \hline
38	&	EPI185840	&	USA 	&	1/6/2009	&	A/Hawaii/02/2009	\\ \hline
39	&	EPI185842	&	USA 	&	1/30/2009	&	A/Hawaii/05/2009	\\ \hline
40	&	EPI185845	&	USA 	&	3/28/2009	&	A/Hawaii/06/2009	\\ \hline
41	&	EPI185851	&	USA 	&	3/30/2009	&	A/Hawaii/07/2009	\\ \hline
42	&	EPI185848	&	USA 	&	3/30/2009	&	A/Hawaii/07/2009	\\ \hline
43	&	EPI185854	&	USA 	&	2/11/2009	&	A/Hawaii/10/2009	\\ \hline
44	&	EPI185857	&	USA 	&	4/15/2009	&	A/Hawaii/14/2009	\\ \hline
45	&	EPI185860	&	USA 	&	4/19/2009	&	A/Hawaii/15/2009	\\ \hline
46	&	EPI185863	&	USA 	&	4/20/2009	&	A/Hawaii/16/2009	\\ \hline
47	&	EPI172430	&	USA 	&	11/10/2008	&	A/Hawaii/47/2008	\\ \hline
48	&	EPI185732	&	China	&	2/1/2009	&	A/Heilongjiang-Nangang/134/2009	\\ \hline
49	&	EPI185735	&	China	&	2/10/2009	&	A/Heilongjiang-Xiangfang/163/2009	\\ \hline
50	&	EPI185779	&	Honduras	&	5/5/2009	&	A/Honduras/56/2009	\\ \hline
51	&	EPI185782	&	Honduras	&	5/6/2009	&	A/Honduras/639/2009	\\ \hline
52	&	EPI185738	&	China	&	2009/00/00	&	A/Hong Kong/03/2009	\\ \hline
53	&	EPI185741	&	China	&	4/7/2009	&	A/Hong Kong/1999/2009	\\ \hline
54	&	EPI185744	&	China	&	4/7/2009	&	A/Hong Kong/2000/2009	\\ \hline
55	&	EPI185747	&	China	&	4/6/2009	&	A/Hong Kong/2007/2009	\\ \hline
56	&	EPI185750	&	China	&	4/5/2009	&	A/Hong Kong/2008/2009	\\ \hline
57	&	EPI175154	&	China	&	10/8/2008	&	A/Hong Kong/3134/2008	\\ \hline
58	&	EPI185754	&	China	&	1/6/2009	&	A/Hong Kong/40/2009	\\ \hline
59	&	EPI187329	&	Japan 	&	1/6/2009	&	A/IWATE/4/2009	\\ \hline
60	&	EPI185866	&	USA 	&	3/14/2009	&	A/Idaho/04/2009	\\ \hline
61	&	EPI172414	&	USA 	&	10/7/2008	&	A/Idaho/17/2008	\\ \hline
62	&	EPI171404	&	USA 	&	10/16/2008	&	A/Idaho/18/2008	\\ \hline
63	&	EPI185868	&	USA 	&	1/19/2009	&	A/Indiana/02/2009	\\ \hline
64	&	EPI185871	&	USA 	&	4/15/2009	&	A/Indiana/10/2009	\\ \hline
65	&	EPI185873	&	USA 	&	1/2/2009	&	A/Iowa/03/2009	\\ \hline
66	&	EPI173166	&	Kenya 	&	10/8/2008	&	A/Isiolo/7514/2008	\\ \hline
67	&	EPI185785	&	Jamaica	&	4/28/2009	&	A/Jamaica/2970/2009	\\ \hline
68	&	EPI185757	&	China	&	3/12/2009	&	A/Jiangsu-Quanshan/114/2009	\\ \hline
69	&	EPI187676	&	China	&	2/13/2009	&	A/Jiangsu-quanshan/38/2009	\\ \hline
70	&	EPI187330	&	China	&	2/24/2009	&	A/Jiangxi-xihu/37/2009	\\ \hline
71	&	EPI169726	&	Japan 	&	11/12/2008	&	A/KANAGAWA/61/2008	\\ \hline
72	&	EPI169727	&	Japan 	&	11/12/2008	&	A/KANAGAWA/65/2008	\\ \hline
73	&	EPI187331	&	Japan	&	1/26/2009	&	A/KOBE/33/2009	\\ \hline
74	&	EPI187332	&	Japan	&	12/17/2008	&	A/KOBE/61/2008	\\ \hline
75	&	EPI187333	&	Japan	&	4/5/2009	&	A/KOBE/84/2009	\\ \hline
76	&	EPI187334	&	Japan	&	1/23/2009	&	A/KOCHI/35/2009	\\ \hline
77	&	EPI187335	&	Japan	&	12/6/2008	&	A/KOCHI/53/2008	\\ \hline
78	&	EPI187337	&	Japan 	&	1/23/2009	&	A/KYOTO/12/2009	\\ \hline
79	&	EPI187338	&	Japan 	&	11/25/2008	&	A/KYOTO/19/2008	\\ \hline
80	&	EPI187339	&	Japan 	&	11/19/2008	&	A/KYOTO/20/2008	\\ \hline
81	&	EPI187677	&	Japan 	&	5/22/2009	&	A/KYOTO/30/2009	\\ \hline
82	&	EPI187340	&	Japan 	&	1/16/2009	&	A/KYOTO/8/2009	\\ \hline
83	&	EPI185875	&	USA 	&	2009/00/00	&	A/Kentucky/02/2009	\\ \hline
84	&	EPI185789	&	Kenya	&	10/3/2008	&	A/Kenya/2234/2008	\\ \hline
85	&	EPI185791	&	Kenya	&	10/15/2008	&	A/Kenya/2282/2008	\\ \hline
86	&	EPI185793	&	Kenya	&	10/21/2008	&	A/Kenya/2297/2008	\\ \hline
87	&	EPI173167	&	Kenya	&	11/7/2008	&	A/Kericho/7532/2008	\\ \hline
88	&	EPI173168	&	Kenya	&	11/10/2008	&	A/Kericho/7533/2008	\\ \hline
89	&	EPI173169	&	Kenya	&	11/10/2008	&	A/Kericho/7534/2008	\\ \hline
90	&	EPI173173	&	Kenya 	&	10/6/2008	&	A/Kisii/7544/2008	\\ \hline
91	&	EPI173174	&	Kenya 	&	10/8/2008	&	A/Kisii/7546/2008	\\ \hline
92	&	EPI173175	&	Kenya 	&	10/14/2008	&	A/Kisii/7548/2008	\\ \hline
93	&	EPI173176	&	Kenya 	&	10/15/2008	&	A/Kisii/7549/2008	\\ \hline
94	&	EPI173177	&	Kenya 	&	10/16/2008	&	A/Kisii/7551/2008	\\ \hline
95	&	EPI173178	&	Kenya 	&	10/17/2008	&	A/Kisii/7552/2008	\\ \hline
96	&	EPI173179	&	Kenya 	&	10/17/2008	&	A/Kisii/7553/2008	\\ \hline
97	&	EPI173180	&	Kenya 	&	10/17/2008	&	A/Kisii/7555/2008	\\ \hline
98	&	EPI173181	&	Kenya 	&	10/21/2008	&	A/Kisii/7556/2008	\\ \hline
99	&	EPI173182	&	Kenya 	&	10/23/2008	&	A/Kisii/7558/2008	\\ \hline
100	&	EPI173183	&	Kenya 	&	10/28/2008	&	A/Kisii/7563/2008	\\ \hline
101	&	EPI173184	&	Kenya 	&	11/4/2008	&	A/Kisii/7566/2008	\\ \hline
102	&	EPI173185	&	Kenya 	&	11/5/2008	&	A/Kisii/7568/2008	\\ \hline
103	&	EPI173186	&	Kenya 	&	11/5/2008	&	A/Kisii/7569/2008	\\ \hline
104	&	EPI173193	&	Kenya 	&	10/3/2008	&	A/Kisumu/7599/2008	\\ \hline
105	&	EPI173194	&	Kenya 	&	10/6/2008	&	A/Kisumu/7600/2008	\\ \hline
106	&	EPI173195	&	Kenya 	&	10/6/2008	&	A/Kisumu/7601/2008	\\ \hline
107	&	EPI173196	&	Kenya 	&	10/6/2008	&	A/Kisumu/7602/2008	\\ \hline
108	&	EPI173197	&	Kenya 	&	10/7/2008	&	A/Kisumu/7603/2008	\\ \hline
109	&	EPI173198	&	Kenya 	&	10/7/2008	&	A/Kisumu/7604/2008	\\ \hline
110	&	EPI173199	&	Kenya 	&	10/8/2008	&	A/Kisumu/7605/2008	\\ \hline
111	&	EPI173200	&	Kenya 	&	10/21/2008	&	A/Kisumu/7609/2008	\\ \hline
112	&	EPI173201	&	Kenya 	&	10/24/2008	&	A/Kisumu/7612/2008	\\ \hline
113	&	EPI173202	&	Kenya 	&	10/28/2008	&	A/Kisumu/7613/2008	\\ \hline
114	&	EPI173203	&	Kenya 	&	10/29/2008	&	A/Kisumu/7614/2008	\\ \hline
115	&	EPI173204	&	Kenya 	&	11/4/2008	&	A/Kisumu/7618/2008	\\ \hline
116	&	EPI173205	&	Kenya 	&	11/4/2008	&	A/Kisumu/7621/2008	\\ \hline
117	&	EPI173164	&	Kenya 	&	11/10/2008	&	A/Kisumu/7627/2008	\\ \hline
118	&	EPI185811	&	South Korea	&	12/8/2008	&	A/Korea/7719/2008	\\ \hline
119	&	EPI185813	&	South Korea	&	12/9/2008	&	A/Korea/7723/2008	\\ \hline
120	&	EPI185815	&	South Korea	&	12/13/2008	&	A/Korea/7858/2008	\\ \hline
121	&	EPI185818	&	South Korea	&	12/13/2008	&	A/Korea/7858/2008	\\ \hline
122	&	EPI185797	&	Kuwait	&	1/18/2009	&	A/Kuwait/494/2009	\\ \hline
123	&	EPI185760	&	China	&	12/4/2008	&	A/Liaoning-Tiedong/14/2008	\\ \hline
124	&	EPI169728	&	Japan 	&	12/3/2008	&	A/MIE/34/2008	\\ \hline
125	&	EPI169729	&	Japan 	&	12/3/2008	&	A/MIE/36/2008	\\ \hline
126	&	EPI187341	&	Japan 	&	12/22/2008	&	A/MIE/43/2008	\\ \hline
127	&	EPI187343	&	Japan 	&	12/3/2008	&	A/MIYAGI/36/2008	\\ \hline
128	&	EPI186304	&	Malaysia	&	1/6/2009	&	A/Malaysia/0032/2009	\\ \hline
129	&	EPI173187	&	Kenya 	&	10/2/2008	&	A/Malindi/7579/2008	\\ \hline
130	&	EPI173188	&	Kenya 	&	10/9/2008	&	A/Malindi/7581/2008	\\ \hline
131	&	EPI173189	&	Kenya 	&	11/11/2008	&	A/Malindi/7590/2008	\\ \hline
132	&	EPI185877	&	USA 	&	1/20/2009	&	A/Maryland/02/2009	\\ \hline
133	&	EPI175206	&	USA 	&	12/5/2008	&	A/Maryland/10/2008	\\ \hline
134	&	EPI185879	&	USA 	&	1/20/2009	&	A/Massachusetts/01/2009	\\ \hline
135	&	EPI185881	&	USA 	&	1/9/2009	&	A/Massachusetts/02/2009	\\ \hline
136	&	EPI171405	&	USA 	&	10/29/2008	&	A/Massachusetts/06/2008	\\ \hline
137	&	EPI172416	&	USA 	&	10/29/2008	&	A/Massachusetts/06/2008	\\ \hline
138	&	EPI175211	&	USA 	&	12/12/2008	&	A/Massachusetts/13/2008	\\ \hline
139	&	EPI173190	&	Kenya 	&	11/12/2008	&	A/Mbagathi/7593/2008	\\ \hline
140	&	EPI185800	&	Mexico	&	3/11/2009	&	A/Mexico/2779/2009	\\ \hline
141	&	EPI175226	&	USA 	&	12/28/2008	&	A/Minnesota/36/2008	\\ \hline
142	&	EPI185883	&	USA 	&	12/23/2008	&	A/Minnesota/37/2008	\\ \hline
143	&	EPI172444	&	USA 	&	12/9/2008	&	A/Montana/02/2008	\\ \hline
144	&	EPI185886	&	USA 	&	3/3/2009	&	A/Montana/04/2009	\\ \hline
145	&	EPI185889	&	USA 	&	1/26/2009	&	A/Montana/05/2009	\\ \hline
146	&	EPI169730	&	Japan 	&	11/20/2008	&	A/NAGANO/1201/2008	\\ \hline
147	&	EPI187353	&	Japan	&	12/18/2008	&	A/NAGOYA/10/2008	\\ \hline
148	&	EPI187357	&	Japan	&	2009/00/00	&	A/NAGOYA/6/2009	\\ \hline
149	&	EPI187359	&	Japan 	&	3/12/2009	&	A/NIIGATA/403/2009	\\ \hline
150	&	EPI187679	&	Japan 	&	5/19/2009	&	A/NIIGATA/603/2009	\\ \hline
151	&	EPI187680	&	Japan 	&	6/1/2009	&	A/NIIGATA/662/2009	\\ \hline
152	&	EPI185891	&	USA 	&	12/22/2008	&	A/Nebraska/08/2008	\\ \hline
153	&	EPI185894	&	USA 	&	4/8/2009	&	A/Nevada/06/2009	\\ \hline
154	&	EPI185896	&	USA 	&	2/16/2009	&	A/New Hampshire/01/2009	\\ \hline
155	&	EPI185899	&	USA 	&	2/28/2009	&	A/New Hampshire/04/2009	\\ \hline
156	&	EPI185900	&	USA 	&	1/27/2009	&	A/New Mexico/03/2009	\\ \hline
157	&	EPI185904	&	USA 	&	2/22/2009	&	A/New York/67/2009	\\ \hline
158	&	EPI185906	&	USA 	&	12/29/2008	&	A/North Carolina/08/2008	\\ \hline
159	&	EPI175252	&	USA 	&	12/15/2008	&	A/North Dakota/02/2008	\\ \hline
160	&	EPI187682	&	Japan 	&	5/20/2009	&	A/OKINAWA/25/2009	\\ \hline
161	&	EPI187684	&	Japan 	&	5/22/2009	&	A/OKINAWA/38/2009	\\ \hline
162	&	EPI185908	&	USA 	&	2/24/2009	&	A/Oregon/03/2009	\\ \hline
163	&	EPI185910	&	USA 	&	1/22/2009	&	A/Pennsylvania/02/2009	\\ \hline
164	&	EPI185913	&	USA 	&	3/8/2009	&	A/Pennsylvania/05/2009	\\ \hline
165	&	EPI175262	&	USA 	&	12/11/2008	&	A/Pennsylvania/13/2008	\\ \hline
166	&	EPI187703	&	Australia	&	4/7/2009	&	A/Perth/15/2009	\\ \hline
167	&	EPI187702	&	Australia	&	4/7/2009	&	A/Perth/15/2009	\\ \hline
168	&	EPI185159	&	Australia	&	4/7/2009	&	A/Perth/15/2009	\\ \hline
169	&	EPI182941	&	Australia	&	4/7/2009	&	A/Perth/16/2009	\\ \hline
170	&	EPI185157	&	Australia	&	4/7/2009	&	A/Perth/16/2009	\\ \hline
171	&	EPI182939	&	Philippines	&	3/5/2009	&	A/Philippines/16/2009	\\ \hline
172	&	EPI173240	&	Philippines	&	11/26/2008	&	A/Philippines/3803/2008	\\ \hline
173	&	EPI182937	&	Philippines	&	3/5/2009	&	A/Philippines/5/2009	\\ \hline
174	&	EPI185802	&	Puerto Rico	&	2/9/2009	&	A/Puerto Rico/18/2009	\\ \hline
175	&	EPI185709	&	Canada 	&	1/23/2009	&	A/Quebec/RV0623/2009	\\ \hline
176	&	EPI185712	&	Canada 	&	2/11/2009	&	A/Quebec/RV0899/2009	\\ \hline
177	&	EPI185715	&	Canada 	&	3/3/2009	&	A/Quebec/RV1069/2009	\\ \hline
178	&	EPI185718	&	Canada 	&	3/3/2009	&	A/Quebec/RV1069/2009	\\ \hline
179	&	EPI187360	&	Japan 	&	12/19/2008	&	A/SAGA/204/2008	\\ \hline
180	&	EPI187364	&	Japan 	&	12/19/2008	&	A/SAITAMA/47/2008	\\ \hline
181	&	EPI169732	&	Japan	&	11/9/2008	&	A/SAKAI/33/2008	\\ \hline
182	&	EPI187685	&	Japan	&	4/28/2009	&	A/SAKAI/41/2009	\\ \hline
183	&	EPI187366	&	Japan	&	12/6/2008	&	A/SAKAI/53/2008	\\ \hline
184	&	EPI187686	&	Japan	&	5/13/2009	&	A/SAPPORO/112/2009	\\ \hline
185	&	EPI172133	&	Japan 	&	12/9/2008	&	A/SHIMANE/106/2008	\\ \hline
186	&	EPI172134	&	Japan 	&	12/8/2008	&	A/SHIMANE/107/2008	\\ \hline
187	&	EPI172135	&	Japan	&	12/18/2008	&	A/SHIZUOKA-C/55/2008	\\ \hline
188	&	EPI187374	&	Japan	&	12/26/2008	&	A/SHIZUOKA-C/57/2008	\\ \hline
189	&	EPI187377	&	Japan 	&	2/20/2009	&	A/SHIZUOKA/428/2009	\\ \hline
190	&	EPI187378	&	Japan 	&	2/23/2009	&	A/SHIZUOKA/448/2009	\\ \hline
191	&	EPI187379	&	Japan 	&	2/26/2009	&	A/SHIZUOKA/502/2009	\\ \hline
192	&	EPI187687	&	Japan 	&	5/23/2009	&	A/SHIZUOKA/736/2009	\\ \hline
193	&	EPI187688	&	Japan 	&	6/14/2009	&	A/SHIZUOKA/791/2009	\\ \hline
194	&	EPI185720	&	Canada 	&	1/5/2009	&	A/Saskatchewan/RV0101/2009	\\ \hline
195	&	EPI185763	&	China	&	2/22/2009	&	A/Shanghai-Nanhui/146/2009	\\ \hline
196	&	EPI185766	&	China	&	3/12/2009	&	A/Shanghai-Nanhui/190/2009	\\ \hline
197	&	EPI185769	&	China	&	11/4/2008	&	A/Sichuan-Qingyang/1396/2008	\\ \hline
198	&	EPI175268	&	China 	&	11/4/2008	&	A/Sichuan/Qingyang1396/2008	\\ \hline
199	&	EPI190150	&	Singapore	&	6/1/2009	&	A/Singapore/33/2009	\\ \hline
200	&	EPI182943	&	Singapore	&	3/27/2009	&	A/Singapore/39/2009	\\ \hline
201	&	EPI173242	&	Singapore	&	11/25/2008	&	A/Singapore/619/2008	\\ \hline
202	&	EPI185915	&	USA 	&	12/31/2008	&	A/South Carolina/07/2008	\\ \hline
203	&	EPI185808	&	Russia 	&	3/30/2009	&	A/St. Petersburg/92/2009	\\ \hline
204	&	EPI186310	&	Australia 	&	1/12/2009	&	A/Sydney/1/2009	\\ \hline
205	&	EPI186311	&	Australia 	&	1/20/2009	&	A/Sydney/3/2009	\\ \hline
206	&	EPI187699	&	Japan 	&	5/18/2009	&	A/TOCHIGI/140/2009	\\ \hline
207	&	EPI169744	&	Japan 	&	11/4/2008	&	A/TOKUSHIMA/18/2008	\\ \hline
208	&	EPI187394	&	Japan 	&	12/18/2008	&	A/TOTTORI/82/2008	\\ \hline
209	&	EPI187694	&	Taiwan	&	2/13/2009	&	A/Taiwan/240/2009	\\ \hline
210	&	EPI169737	&	Taiwan	&	10/3/2008	&	A/Taiwan/540/2008	\\ \hline
211	&	EPI169738	&	Taiwan	&	10/3/2008	&	A/Taiwan/542/2008	\\ \hline
212	&	EPI169739	&	Taiwan	&	10/12/2008	&	A/Taiwan/548/2008	\\ \hline
213	&	EPI169740	&	Taiwan	&	10/18/2008	&	A/Taiwan/549/2008	\\ \hline
214	&	EPI169741	&	Taiwan	&	10/31/2008	&	A/Taiwan/573/2008	\\ \hline
215	&	EPI169742	&	Taiwan	&	11/7/2008	&	A/Taiwan/580/2008	\\ \hline
216	&	EPI169743	&	Taiwan	&	11/13/2008	&	A/Taiwan/599/2008	\\ \hline
217	&	EPI172448	&	Taiwan	&	10/25/2008	&	A/Taiwan/611/2008	\\ \hline
218	&	EPI172450	&	Taiwan	&	11/8/2008	&	A/Taiwan/612/2008	\\ \hline
219	&	EPI187696	&	Taiwan	&	5/2/2009	&	A/Taiwan/712/2009	\\ \hline
220	&	EPI187697	&	Taiwan	&	5/2/2009	&	A/Taiwan/714/2009	\\ \hline
221	&	EPI187698	&	Taiwan	&	4/23/2009	&	A/Taiwan/725/2009	\\ \hline
222	&	EPI185918	&	USA 	&	1/26/2009	&	A/Texas/02/2009	\\ \hline
223	&	EPI185920	&	USA 	&	3/4/2009	&	A/Texas/03/2009	\\ \hline
224	&	EPI185923	&	USA 	&	3/18/2009	&	A/Texas/18/2009	\\ \hline
225	&	EPI175295	&	USA 	&	12/26/2008	&	A/Texas/39/2008	\\ \hline
226	&	EPI185820	&	Thailand	&	10/3/2008	&	A/Thailand/860/2008	\\ \hline
227	&	EPI175300	&	Thailand	&	11/6/2008	&	A/Thailand/944/2008	\\ \hline
228	&	EPI175302	&	Thailand	&	11/3/2008	&	A/Thailand/947/2008	\\ \hline
229	&	EPI185823	&	Thailand	&	11/12/2008	&	A/Thailand/981/2008	\\ \hline
230	&	EPI175305	&	Thailand	&	11/3/2008	&	A/Thailand/998/2008	\\ \hline
231	&	EPI185772	&	China	&	3/3/2009	&	A/Tianjin-Hongqiao/141/2009	\\ \hline
232	&	EPI185774	&	China	&	12/30/2008	&	A/Tianjin-Nankai/1211/2008	\\ \hline
233	&	EPI185949	&	Trinidad and Tobago	&	2009/00/00	&	A/Trinidad/2940/2009	\\ \hline
234	&	EPI185925	&	USA 	&	2/13/2009	&	A/Vermont/02/2009	\\ \hline
235	&	EPI189220	&	Australia 	&	6/2/2009	&	A/Victoria/208/2009	\\ \hline
236	&	EPI189218	&	Australia 	&	6/2/2009	&	A/Victoria/209/2009	\\ \hline
237	&	EPI190148	&	Australia 	&	6/2/2009	&	A/Victoria/210/2009	\\ \hline
238	&	EPI173254	&	Australia 	&	1/12/2009	&	A/Victoria/500/2009	\\ \hline
239	&	EPI176481	&	Australia 	&	1/25/2009	&	A/Victoria/502/2009	\\ \hline
240	&	EPI185927	&	USA 	&	1/5/2009	&	A/Virginia/02/2009	\\ \hline
241	&	EPI187395	&	Japan 	&	1/13/2009	&	A/WAKAYAMA/14/2009	\\ \hline
242	&	EPI185929	&	USA 	&	1/25/2009	&	A/Washington/04/2009	\\ \hline
243	&	EPI185931	&	USA 	&	3/3/2009	&	A/Washington/05/2009	\\ \hline
244	&	EPI185934	&	USA 	&	3/30/2009	&	A/Washington/06/2009	\\ \hline
245	&	EPI172418	&	USA 	&	12/8/2008	&	A/Washington/09/2008	\\ \hline
246	&	EPI185937	&	USA 	&	4/6/2009	&	A/Washington/15/2009	\\ \hline
247	&	EPI185940	&	USA 	&	3/16/2009	&	A/Washington/16/2009	\\ \hline
248	&	EPI185942	&	USA 	&	1/7/2009	&	A/Wisconsin/03/2009	\\ \hline
249	&	EPI185944	&	USA 	&	2/5/2009	&	A/Wisconsin/05/2009	\\ \hline
250	&	EPI172434	&	USA 	&	12/3/2008	&	A/Wisconsin/18/2008	\\ \hline
251	&	EPI175344	&	USA 	&	12/25/2008	&	A/Wisconsin/23/2008	\\ \hline
252	&	EPI175346	&	USA 	&	12/23/2008	&	A/Wisconsin/24/2008	\\ \hline
253	&	EPI185946	&	USA 	&	2/2/2009	&	A/Wyoming/02/2009	\\ \hline
254	&	EPI187396	&	Japan 	&	11/15/2008	&	A/YAMAGATA/111/2008	\\ \hline
255	&	EPI187397	&	Japan 	&	11/16/2008	&	A/YAMAGATA/112/2008	\\ \hline
256	&	EPI187398	&	Japan 	&	1/13/2009	&	A/YAMAGUCHI/15/2009	\\ \hline
257	&	EPI172138	&	Japan 	&	12/3/2008	&	A/YAMAGUCHI/30/2008	\\ \hline
258	&	EPI172139	&	Japan 	&	12/8/2008	&	A/YAMAGUCHI/35/2008	\\ \hline
259	&	EPI187400	&	Japan 	&	11/5/2008	&	A/YAMANASHI/140/2008	\\ \hline
260	&	EPI187402	&	Japan	&	3/8/2009	&	A/YOKOHAMA/107/2009	\\ \hline
261	&	EPI187403	&	Japan	&	3/4/2009	&	A/YOKOHAMA/108/2009	\\ \hline
262	&	EPI187404	&	Japan	&	1/27/2009	&	A/YOKOHAMA/52/2009	\\ \hline
263	&	EPI172142	&	Japan	&	12/2/2008	&	A/YOKOHAMA/97/2008	\\ \hline

\label{sequences2}
\end{longtable}
